\documentclass[]{aa}

\usepackage{txfonts,graphicx,rotating,natbib}
\bibpunct{(}{)}{;}{a}{}{,}

\newcommand{\oder}[2]{{{\rm d}\,#1\over{\rm d}\,#2}}
\newcommand{\pd}[2]{\,{\partial\,#1\over\partial\,#2}\,}
\newcommand{\mps}{Meteoritics~\&~Plan.~Sci.}%
\newcommand{\areps}{Ann.~Rev.~Earth~\&~Plan.~Sci.}%
\newcommand{\epsl}{Earth~\&~Plan.~Sci.~Lett.}%
\newcommand{\lpscl}{Lunar Planet. Sci. Conf. Lett.}%
\newcommand{\tectp}{Tectonophysics}%
\newcommand{\jap}{J.~Appl.~Phys.}%
\newcommand{\jacs}{J.~Americ.~Ceramic~Soc.}%

\begin{document}

\title{Thermal evolution and sintering of chondritic planetesimals}

\author{Stephan Henke\inst{1} \and Hans-Peter Gail\inst{1}
 \and Mario Trieloff\inst{2} \and Winfried H. Schwarz\inst{2}
 \and Thorsten Kleine\inst{3}
}

\institute{
Institut f\"ur Theoretische Astrophysik, Zentrum f\"ur Astronomie, 
           Universit\"at Heidelberg, Albert-\"Uberle-Str. 2,
           69120 Heidelberg, Germany 
\and
Institut f\"ur Geowissenschaften, Universit\"at Heidelberg, Im Neuenheimer
           Feld 236, 69120 Heidelberg, Germany
\and
Institut f\"ur Planetologie, Universit\"at M\"unster, Wilhelm-Klemm-Str. 10,
            48149 M\"unster, Germany, 
  }

\offprints{\tt gail@ita.uni-heidelberg.de}

\date{Received date ; accepted date}

\abstract
{}
{Radiometric ages for chondritic meteorites and their components provide information on the accretion timescale of chondrite parent bodies, and on cooling paths within certain areas of these bodies. However, to utilize this age information for constraining the internal structure, and the accretion and cooling history of the chondrite parent bodies, the empirical cooling paths obtained by dating chondrites must be combined with theoretical models of the thermal evolution of planetesimals. Important parameters in such thermal models include the initial abundances of heat-producing short-lived radionuclides ($^{26\!}$Al and $^{60}$Fe), which are determined by the accretion timescale, and the terminal size, chemical composition and physical properties of the chondritic planetesimals. The major aim of this study is to assess the effects of sintering of initially porous material on the thermal evolution of planetesimals, and to constrain the values of basic parameters that determined the structure and evolution of the H chondrite parent body. 
}
{A new code is presented for modelling the thermal evolution of ordinary chondrite parent bodies that initially are highly porous and undergo sintering by hot pressing as they are heated by decay of radioactive nuclei. The pressure and temperature stratification in the interior of the bodies is calculated by solving the equations of hydrostatic equilibrium and energy transport. The decrease of porosity of the granular material by hot pressing due to self-gravity is followed by solving a set of equations for the sintering of powder materials. For the heat conductivity of granular material we combine recently measured data for highly porous powder materials, relevant for the surface layers of planetesimals, with data for heat conductivity of chondrite material, relevant for the strongly sintered material in deeper layers.
}
{Our new model demonstrates that in initially porous planetesimals heating to
central temperatures sufficient for melting can occur for bodies a few km in
size, that is, a factor of $\approx10$ smaller than for compact bodies.
Furthermore, for high initial $^{60}$Fe abundances small bodies may
differentiate even when they had formed as late as 3-4 Ma after CAI formation.
To demonstrate the capability of our new model, the thermal evolution of the H
chondrite parent body was reconstructed. The model starts with a porous body
that is later compacted first by `cold pressing' at low temperatures and then by
`hot pressing' for temperatures above $\approx700$\,K, i.e., the threshold
temperature for sintering of silicates. The thermal model was fitted to the well
constrained cooling histories of the two H chondrites Kernouv\'e (H6) and
Richardton (H5). The best fit is obtained for a parent body with a radius of 
100\,km that accreted at $t=2.3$\,Ma after CAI formation, and had an initial 
$^{60}$Fe/$^{56}$Fe $=4.1\times10^{-7}$. Burial depths of 8.3\,km and 36\,km for
Richardton and Kernouv\'e can reproduce their empirically determined cooling
history. These burial depths are shallower than those derived in previous
models. This reflects the strong insulating effect of the residual powder
surface layer, which is characterized by a low thermal conductivity.}
{}

\keywords{Solar system: formation, planetary systems: formation, 
planetary systems: protoplanetary disks}

\maketitle

\section{%
Introduction}

According to our present understanding of the formation of terrestrial planets,
planetesimals from the size class of 100\,km form an important transition state
during the formation of protoplanets and planets \citep[e.g.][]{Wei06,Naga07}.
These bodies are sufficiently big that heat liberated by decay of short lived
radioactive nuclei, e.g. $^{26}$Al, does not easily flow to the surface and 
radiates away. Instead they heat up and the bigger ones start to melt in
their core region. The pristine dust material from the accretion disc undergoes
in  this way a number of metamorphic processes until it evolves into planetary
material. A few such bodies representing this intermediate stage of the 
planetary formation process have survived in the asteroid belt and material
from such bodies is available on earth as meteorites. Meteorites preserve in
their structure and composition rather detailed information on the processes
that were at work during planetary formation. Recovering this information from
these witnesses of the early history of our solar system requires to model the
structure, composition, and thermal history of meteorite parent bodies in order
to put the analytic results of laboratory investigations of individual
meteorites into a more general context. 
   
There are fundamental differences in the composition of ordinary and
carbonaceous chondritic meteorites that are related to the absence or presence
of ice, respectively, in the region of the protoplanetary disc where their parent
bodies formed. We aim in this paper to study the parent bodies of ordinary
chondrites. They were essentially ice-free, and are in this respect more similar
to planetesimals in the inner solar system, where the terrestrial planets formed.
In particular we concentrate on the H and L chondrites that form two rather
homogeneous classes and seem each to descend from a single parent body in the
asteroid belt.

The model calculations for the thermal evolution of asteroids done so far are
reviewed by \citet{McS03} and by \citet{Gho06}. Many of the calculations are
based on the analytic solution of the heat conduction equation with constant
coefficients for a spherical body heated by a homogeneously distributed and
exponentially decaying heat source ($^{26\!}$Al) given by \citet{Miy81}. Such 
models have the advantage of being simple and easily applicable for discussions
of special problems, but they cannot be extended to more realistic cases where
material properties like heat conduction, specific heats and others are neither
spatially nor temporally constant, or to include additional physics. It was
possible, however, to show that H and L chondrites of petrologic classes 3 to 6
originate from bodies of about 100 km size, heated by radioactive decay of
short lived nuclei, and that the different petrologic classes correspond to
layers at different depths within the same body which experienced different
thermal histories during the initial heating and subsequent cooling of the body
\citep{Goe94,Tri03,Kle08,Har10}. 

If more physics is to be included in a model a numerical solution of the basic
set of equations is necessary. One of the first models of this kind was the
model calculation by \citet{Yom84} that used data for the temperature and
porosity dependence of material properties which they determined by laboratory
measurements on H and L chondrite material \citep{Yom83}. A central point of
this model calculation was that it addressed for the first time quantitatively
the process of sintering of the initially strongly porous material under the
action of pressure resulting from self-gravity once the body is heated to high
temperatures (\,$\gtrsim700$\,K). As a result, the body develops a strong zoning
of the material properties: A highly consolidated core with high thermal
conductivity if temperature and pressure become sufficiently high during the
course of the thermal evolution of the body, and a porous outer layer with low
heat conductivity where temperatures and pressures never climbed to the level
where compaction by sintering becomes possible.  

This model shows distinct basic differences to a model with constant coefficients
like that of \citet{Miy81} since it has an extended inner region with an only
slow outwards drop of temperature where heat conductivity is high, and a rather
shallow outer layer where temperature rapidly drops to the outer boundary
temperature. The results for the depths of formation of different petrologic
types and the predicted spatial extent of such zones are very different in the
models of \citet{Miy81} and \citet{Yom84}. This demonstrates the importance 
of including the consolidation of granular material by 'hot pressing' and the
resulting changes of thermal conductivity in the models. Unfortunately the
results of \citet{Yom84} are somewhat impaired by the fact that the important
heating source $^{26\!}$Al was not included in the calculation. In the sequel
the sintering processes seem to have been included in the modelling
of  thermal evolution of asteroids only in the calculations of \citet{Akr97}
and of \citet{Sen04}. Other studies of the blanketing effect of porous outer
layers on thermal evolution \citep{Akr98,Hev06,Sah07,Gup10} \emph{assume} the
consolidation of such layers at some characteristic temperature, but do not
include this process as part of the model calculation.

In this paper we study the thermal history of the parent bodies of ordinary
chondrites using new data for the thermal conductivity of granular material
\citep{Kra11,Kra11b} and on the compaction by cold isostatic pressing before the onset
of sintering \citep{Gue09}. The modelling of the sintering process is based on
the same kind of theory \citep{Rao72} as in \citet{Yom84}. This theory does
not  correspond to the more recent approaches used for modelling of hot pressing
in technical processes \citep[e.g.][]{Arz83,Fis83,Lar96,Sto99}, but it seems
more appropriate for the lower pressure regime encountered in asteroids.

The main purpose of this paper is to develop improved models for the thermal
history of parent bodies of ordinary chondrites and to compare the model
results with cooling histories of H and L chondrites of different petrologic
types determined from thermochronological methods in cosmochemistry. This 
allows to better constrain the size of the parent bodies and their formation 
times.
 
\begin{table}
\caption{Basic mineral species considered for calculation of properties of
chondrite material, their atomic weight $A$, mass-density $\varrho$, and
abbreviation of mineral name.}

\begin{tabular}{llrll}
\hline
\noalign{\smallskip}
species & composition & \multicolumn{1}{c}{$A$} & 
   \multicolumn{1}{c}{$\varrho$} & Abbr. \\
\noalign{\smallskip}
\hline
\\
Forsterite   & Mg$_2$SiO$_4$       & 140.69 & 3.22 & Fo \\
Fayalite     & Fe$_2$SiO$_4$       & 203.78 & 4.66 & Fa \\
Enstatite    & MgSiO$_3$           & 100.39 & 3.20 & En \\
Ferrosilite  & FeSiO$_3$           & 132.32 & 3.52 & Fs \\
Wollastonite & CaSiO$_2$           & 116.16 & 2.91 & Wo \\
Anorthite    & CaAl$_2$Si$_2$O$_8$ & 277.41 & 2.75 & An \\
Albite       & NaAlSi$_3$O$_8$     & 263.02 & 2.63 & Ab \\
Orthoclase   & KAlSi$_3$O$_8$      & 278.33 & 2.55 & Or \\
Iron         & Fe                  &  55.45 & 7.81 & Irn \\
Nickel       & Ni                  &  58.69 & 8.91 & Nkl \\
Troilite     & FeS                 &  87.91 & 4.91 & Tr \\
\noalign{\smallskip}
\hline
\end{tabular}

\label{TabMinProp}
\end{table}


\section{%
Thermal history of chondrite parent bodies}

The heat source for early differentiation and metamorphism of meteorite parent
bodies was decay heat of short-lived nuclides like $^{26\!}$Al or $^{60}$Fe 
\citep{Goe94,Tri03,Kle08,Bou07}. If planetesimals are larger than tens of km,
the maximum degree of internal heating is given by
the initial $^{26\!}$Al abundance, i.e., mainly formation time. Planetesimals
formed at $\approx$\,2 Ma after calcium-aluminium rich-inclusions (CAIs 
--- commonly regarded as oldest objects in the solar system), heat up without
differentiation, yielding thermally metamorphosed chondritic parent bodies
(ordinary chondrites, enstatite chondrites, or more strongly heated
Acapulcoites and Lodranites). Primitive chondrites (such as petrologic type 3)
can survive in the outer cool layers of larger parent bodies \citep[following 
the onion shell model, see, e.g.,][]{Goe94,Tri03}, or in bodies that never
grew larger than 10 -- 20 km in size \citep{Hev06,Yom84}, or in bodies that
formed relatively late.

\begin{table*} 
\caption{Typical mineral composition of chondrites, mass-densities $\varrho$ of
components, mass-fractions $X_{\rm min}$ of minerals, and mass-fractions 
$X_{\rm el}$ of the elements that release heat by radioactive decays
\citep [data for $X_{\rm min}$ from][]{Yom83}.
}

\centerline{
\begin{tabular}{l@{\hspace{.7cm}}lcc@{\hspace{.7cm}}lcc@{\hspace{1.2cm}}ccc}
\hline
\noalign{\smallskip}
        & \multicolumn{3}{c}{H-chondrite} & \multicolumn{3}{c}{L-chondrite} &
  & H-chondrite & L-chondrite \\ 
species & composition & $\varrho$ & $X_{\rm min}$ & composition & $\varrho$ & 
$X_{\rm min}$ & element & $X_{\rm el}$ & $X_{\rm el}$\\
        &             & g\,cm$^{-3}$ &  &             & g\,cm$^{-3}$ \\
\noalign{\smallskip}
\hline
\\
Olivine       & Fo$_{80}$Fa$_{20}$ & 3.51 & 0.37 & Fo$_{75}$Fa$_{25}$ & 
  3.58 & 0.49 &
   Al & $9.10\times10^{-3}$ & $8.95\times10^{-3}$ \\
Orthopyroxene & En$_{83}$Fs$_{17}$ & 3.25 & 0.25 & En$_{78}$Fs$_{22}$ & 
  3.27 & 0.23 &
   Fe & $2.93\times10^{-1}$ & $2.26\times10^{-1}$ \\
Clinopyroxene & En$_{49}$Fs$_6$Wo$_{45}$ & 3.09 & 0.05 & 
  En$_{48}$Fs$_8$Wo$_{44}$  & 3.10 & 0.06 &
  K & $7.07\times10^{-4}$ & $7.07\times10^{-4}$ \\
Plagioclase &  Ab$_{82}$Or$_6$An$_{12}$ & 2.64 & 0.08 &
  Ab$_{84}$Or$_6$An$_{10}$ & 2.64 & 0.08 &
  Th & $5.16\times10^{-8}$ & $5.72\times10^{-8}$ \\
Nickel-iron & Irn$_{92}$Nkl$_{8}$ & 7.90 & 0.20 & 
  Irn$_{87}$Nkl$_{13}$ & 7.95 & 0.09 &
  U & $2.86\times10^{-8}$ & $3.17\times10^{-8}$ \\
Troilite & Tr & 4.91 & 0.05 & Tr & 4.91 & 0.05 \\
\\
\multicolumn{2}{l}{Average $=\varrho_{\rm bulk}$}  & 3.78 & & & 3.59 & \\ 
\noalign{\smallskip}
\hline
\end{tabular}
 }

\label{TabMinMix}
\end{table*}

Such scenarios can be verified via their thermal history and cooling paths.
These paths can be evaluated by a variety of thermochronological methods
applying  chronometers with different closure temperatures, which means the
temperature where no parent or daughter nuclide is lost from their host
minerals. 
The Hf-W dating method \citep[e.g.][]{Kle05,Kle08} is useful for cooling below
1050 to 1150 K. The closure temperature for the 
U-Pb-Pb system in phosphates is $\approx$\,720 K \citep[e.g.][]{Goe94,Bou07}, 
Ar-Ar ages reflect the cooling below $\approx$\,550 K for oligoclase feldspar. The
annealing temperature for $^{244}$Pu fission tracks \citep[e.g.][]{Pel97,Tri03}
in orthopyroxene is 550 K, for the phosphate merrillite $\approx$\,390 K. The latter
yield a relative cooling rate between 550 and 390 K for the respective rock
\citep[e.g.][]{Pel97,Tri03}. Another method to determine cooling rates between
800 and 600 K are metallographic cooling rates, which use Fe-Ni diffusion
profiles in metal grains consisting of the two different  Fe/Ni phases kamacite 
and taenite \citep[e.g.]{Her94, Hop01}. These data are the basis for the 
simulation of the cooling of chondritic parent bodies in 
Sect.~\ref{SectFitHChond}.

\section{%
Composition of planetesimal material}

\subsection{%
Mineral composition of planetesimal material}
\label{SectMinComp}

From the pristine dust material in the accretion disk, that is formed
simultaneously with the formation of a new star, finally planetesimals are
formed by a complicated agglomeration process. This process is not modelled in
our calculation \citep[for a review on this part of the problem see, e.g.,][]
{Naga07}. We concentrate here on the evolution of the internal constitution of
the planetesimals, the successors of which can still be found in the asteroid
belt. In particular we concentrate on the class of planetesimals that are the
parent bodies of the H- and  L-chondrites. Their composition is well known from
studies of meteorites. The matrix and chondrules that form the material of
the ordinary chondrites consist of a mixture of minerals that themselves
in most cases are solid solutions of a number of components. We consider an
average composition for the material contained in the parent bodies of the H-
and of the L-chondrites that concentrates on the few main constituents that
represent almost 100\% of the matter. The many more additional compounds with
small abundances found in the material are not important for the bulk 
properties of those materials that determine the constitution of the
planetesimals and their evolution. 

The pure mineral components used in our model for the composition of
planetesimals and their mass-densities are listed in Table~\ref{TabMinProp}.
The mixture of minerals that is assumed to form the planetesimal material
is listed in Table~\ref{TabMinMix}. The composition is given in the notion
where, e.g., Fo$_{80}$Fa$_{20}$ denotes that 80\% of all basic building blocks
of the mineral correspond to forsterite and 20\% to fayalite. The quantity
$X_{\rm min}$ denotes the mass fraction of the component in the mixture. The
composition of the mixture components and the fractions $X_{\rm min}$ are
taken from \citet{Yom83}; the average composition of chondrites given by 
\citet{Jar90} is essentially the same, only the Fe and Al mass-fractions are
slightly less than in \citet{Yom83}.

The density of the solid solutions given in the table is simply calculated
as average of the densities of the pure components, weighted with their
mole fractions in the solid solution. This assumes that non-ideal mixing
effects are too small as that they need to be considered for our calculations.
The average density $\varrho_{\rm bulk}$ of the mixture is calculated from
\begin{equation}
\varrho_{\rm bulk}=\left(\sum_i{X_{{\rm min},i}\over \varrho_i}\right)^{-1}\,.
\end{equation}
This is the density of the consolidated meteoritic material, i.e., without
pores.

Table~\ref{TabMinMix} presents mass-fractions $X_{\rm el}$ of the elements for
which radioactive isotopes of sufficient abundance exist that their energy
release during decay contributes significantly to the heating of the
planetesimals. The element abundances used for the calculation are from
\citet{Lod09}.
 
\begin{table}
\caption{Some properties of chondrules and matrix in H and L chondrites
(data from \citet{Sco97}).}

\begin{tabular}{llll}
\hline
\hline
\noalign{\smallskip}
& & H & L \\
\hline
\\
Chondrules & Vol.\% & 60--80 & 60--80 \\
Chondrule diameter (ave.) & mm& 0.3 & 0.5 \\
Matrix & Vol.\% & 10--15 & 10--15 \\
\noalign{\smallskip}
\hline
\end{tabular}
\label{TabCompChondMet}
\end{table}

\subsection{Granular nature of the material}
\label{SectGranular}

The main constituents present in chondritic meteorites are chondrules and a
matrix of fine dust grains. The typical relative abundance of these constituents
is shown in Table \ref{TabCompChondMet}, they have typical sizes of micrometer
in the case of relatively unprocessed matrix, and typical sizes of millimetres
in  the case of chondrules, with variations specific to each chondrite group 
\citep{Sco97}. Though chondrules may also contain fine grained mineral
assemblages, they entered the meteorite parent body as solidified melt droplets
after they formed by unspecified flash heating events in the solar nebula.

The size of the fine dust grains of the matrix before compaction and sintering of
the meteoritic material is not known. The observed sizes of matrix particles may
be not representative because coarsening may have occurred during heating of the
parent body (Ostwald ripening). Probably the very fine granular units in
interplanetary dust particles (IDPs) give a hint to the pristine size of dust
particles in protoplanetary discs before incorporation into the forming bodies of
the solar system. IDPs seem to represent cometary dust \citep{Bra03} and this
seems to be the least processed dust material handed down from the early Solar
Nebula. The study of \citet{Rie93} shows that the granular units forming IDPs are
sub-micrometer sized, most of them having diameters smaller than $0.5\,\mu$m and
many being nano-meter sized, $\ll0.1\,\mu$, with a small population ranging in
size up to  several~$\mu$m. Typical particle radii then are roughly of the order
of $0.1\,\mu$m \dots\ $0.2\,\mu$m.

It is generally assumed that the planetesimal material initially is very
loosely packed with a high degree of porosity. As porosity $\phi$ we denote
the  \emph{volume fraction not filled} by solid dust material. The
volume not filled by the solids is called the pore space or the pores. The pores
may form an interconnected network of voids and channels at high porosity, 
$\phi\lesssim1$, or isolated voids at low porosity, $\phi\ll1$. The pores may be
empty (vacuum) or may be filled with gas or something else. If there are no
pores, the material is called compact.

The compact solid material has density $\varrho_{\rm bulk}$. This material
is a complex mixture of minerals, that is described in Sect.~\ref{SectMinComp}.
The porous material has density
\begin{equation}
\varrho=\varrho_{\rm bulk}(1-\phi)+\varrho_{\rm p}\phi\,,
\end{equation}
where $\varrho_{\rm p}$  is the density of the material in the pores. If the
pores are filled with gas then $\varrho_{\rm p}$ is small enough that we may
assume $\varrho_{\rm p}=0$. At sufficient low temperatures the pores may
alternatively by filled by water and ice in which case $\varrho_{\rm p}$ may 
not be neglected, but here we do not consider such cases. Then
\begin{equation}
\varrho=\varrho_{\rm bulk}(1-\phi)\,.
\label{EOSPor}
\end{equation}
Besides porosity $\phi$ we also consider the filling factor
\begin{equation}
D=1-\phi\,.
\end{equation}
Then
\begin{equation}
\varrho=\varrho_{\rm bulk}D\,.
\label{EqDensD}
\end{equation}
For this reason $D$ is also called the \emph{relative density} of the porous
material.

The porous material may approximately be described as a packing of spheres.
With respect to the chondrules this is not unrealistic, because they show
nearly spherical shape. For the matrix grains this may be taken as a rough
approximation to describe some basic properties of the packing. For a random
packing of equal sized spheres, \citet{Arz82} found a relation between the
coordination number $Z$, i.e., the average number of contact points of a
particle with its neighbours, with the filling factor or relative density $D$:
\begin{eqnarray}
Z(D) = Z_0 +C \left( \left( \frac{D}{D_0} \right)^{1/3} -1\right)
\label{S_Koordinationszahl}
\end{eqnarray}
Here, $Z_0=7.3$ and $D_0=0.64$. These values refer to the coordination number
of a random dense packing of spheres \citep[cf.][]{Jae92}. The constant
$C=15.5$ is the slope of the radial density function (distribution of centre
distances). The approximation hold up to $D>0.9$. 

For a packing of equal sized spheres it is found that there are two critical
filling factors. One is the random close packing with a filling factor of
$D=0.64$, i.e., a porosity of $\phi=0.36$ \citep{Sco62,Jae92}, and an average
coordination number $Z=7.3$. Its formation requires taping or joggling at the
material. The other one is the loosest close packing that is just stable under
the application of external forces (in the limit of vanishing force), which has
$D=0.56$ or $\phi=0.44$ \citep{Ono90,Jae92} and average coordination number 
$Z\approx6.6$. With respect to chondritic material this
means that volume filling factors of more than $D=0.64$ for chondrules, like
that given in Table  \ref{TabCompChondMet}, require that compaction by
sintering or compaction by crushing of particles due to strong pressure
occurred . 

For volume filling factors of chondrules between the random close packing
($D=0.64$) and the random loose packing ($D=0.56$) some pre-compaction by
applied forces occurred. The high volume filling factors of chondrules in the
material of the chondrites, hence, shows that this material was already
significantly compressed. It is not possible to reconstruct from this the
initial properties of the material in the bodies that contributed to the growth
of the parent bodies. We have to estimate this somehow.


At the basis of the planetesimal formation process there stands the agglomeration
of fine dust grains like that found in IDPs into dust aggregates of increasing
size. Such an agglomerated material from very fine grains that was not subject
to any pressure has porosity as high as $\phi\approx0.8\dots0.9$ 
\citep[e.g.][]{Yan00,Kra11}. The number of contacts of a particle to neighbouring 
particles then is on the average equal to about two, i.e., the fine dust grains
form essentially chain like structures. Such high-porosity material seems to
be preserved in some comets \citep{Blu06}. 

Collisions of dust aggregates  during the growth process of planetesimals leads
to compaction of the material. The experiments of \citet{Wei09} have shown that
the porosity can be reduced to $\phi\approx 0.64$ \citep[or even lower,
see][]{Kot10} by repeated impacts. This porosity is still higher than the
random loose packing and, hence, is not guaranteed to be mechanically stable.
The average coordination number is $Z\approx5$. Lower porosities of $\phi
\lesssim0.4$ were obtained by applying static pressures of more than 10\,bar 
\citep{Gue09}. Since the planetesimals form by repeated growth process by
collisions with other growing planetesimals and impact velocities can be
rather high we assume in the model calculations that the initial
porosity of the material from which the parent bodies of the asteroids formed 
is already compacted to some extent and assume a porosity of the order 
of $\phi=0.6\dots0.5$.

\begin{figure*}

\centerline{
\includegraphics[height=5.5cm]{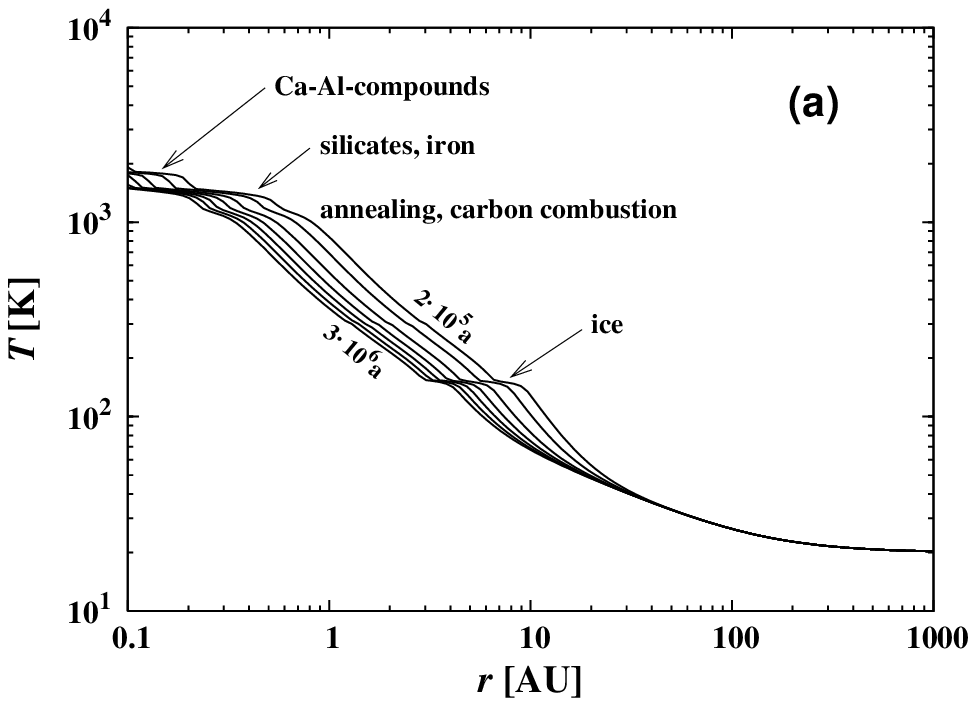}
\hspace{.1cm}
\includegraphics[height=5.5cm]{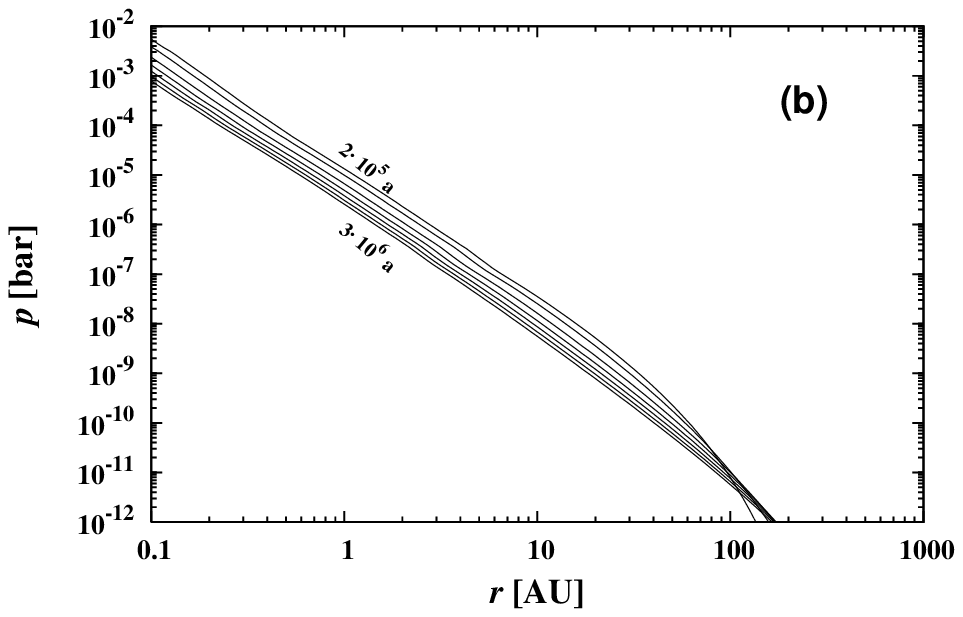}
}

\caption{Variation of {\bf(a)} mid-plane temperature and {\bf(b)} pressure with
distance from the star at different evolutionary epochs (0.2\,Ma and
from 0.5\,Ma to 3\,Ma in steps of  0.5\,Ma) in a model for the evolution of the
Solar Nebula (one-zone $\alpha$-model). They define the outer boundary conditions
for a planetesimal embedded in an accretion disk. In the left part the regions
are indicated where the disappearance of an important absorber results in an
extended region of nearly constant mid-plane temperature.}

\label{FigTempDisk}

\end{figure*}

\section{%
Internal pressure within planetesimals}

\subsection{Hydrostatic equilibrium}

The planetesimals are subject to the mutual gravitational interaction between
their different parts from which there results an inward directed gravitational
force at each location. As long as the material is highly porous
and was not yet subject to sintering processes, the granular material may start to
flow (by rolling and gliding of the powder particles) and evolve into a state with
the densest packing of the granular material. If the planetesimal material is
approximated by a random packing of spheres of equal size its porosity in this
state would be about $\phi=0.37$ \citep{Yan00}. In this state the forces due to
the mutual gravitational attraction of all other particles are compensated
by the reactive forces due to internal stresses that are build up within the 
particles as result of the forces acting between contact points to neighbouring
particles. The material then can exist in a state of hydrostatic equilibrium
with no relative motion between grains. This equilibrium state, however, is of
a rather labile character because at contacts points between grains stresses
may be build up that are strong enough that particles may break apart into
smaller pieces and some further compaction of the material is possible by
application of high pressures. For planetesimal material this kind of compaction
is probably only relevant for impact problems. During the gradual build-up of
planetesimals, temperatures become already high enough by internal heating for
sintering by internal creep to occur before really high pressures are built up
by which the material may be crushed.   

We assume in our further considerations that during the very initial build-up
of planetesimals the initially very loosely packed granular material, under the
action of its own gravitational attraction and/or during collisions with other
planetesimals, has already evolved by granular flow to a state where all
particles have a sufficient number of contacts to neighbours such that further
relative motions are effectively blocked ($\phi\lesssim0.44$). Further
compaction then requires considerable pressures because particles have to be
broken for this. Additionally we neglect that the planetesimals may rotate and
therefore neglect any deformation of the body resulting from this. Then we may
assume that the planetesimals have spherically symmetric structure and the
distribution of internal pressure in the porous dust ball is described by the
well known hydrostatic pressure equation
\begin{equation}
\oder pr=-{GM_r\over r^2}\,\varrho
\label{EqHydro}
\end{equation}
where
\begin{equation}
M_r=4\pi\int_0^r{\rm d}r'\,r'^2\varrho(r')\,.
\label{DefMr}
\end{equation}
The density $\varrho$ is the mass-density of the material. 

As long as a planetesimal is embedded in an accretion disk it is subject to the
external pressure in the disk. The equation of the hydrostatic pressure
stratification in the planetesimal therefore has to be solved with the outer
boundary condition at the outer radius $R$
\begin{equation}
p(R)=p_{\rm ext} \,,
\end{equation} 
with $p_{\rm ext}$ being the pressure in the accretion disk. The pressure in
the midplane of the Solar Nebula at a number of instants as calculated from an
evolution model of the disk \citep{Weh02,Weh08} is shown in 
Fig.~\ref{FigTempDisk}b. This defines the outer boundary condition $p_{\rm ext}$.
After dissipation of the accretion disk one has to use
\begin{equation}
p(R)=0
\end{equation}
instead. 

In order to estimate the magnitude of pressure let us consider a body with
constant density $\varrho$. We have in this case
\begin{equation}
M_r={4\pi\over3}\varrho r^3\,.
\end{equation}
For a body with radius $R$ and external pressure $p_{\rm ext}$ integration
of the pressure equation yields
\begin{equation}
p(r)=p_{\rm ext}+{4\pi\over3}G\varrho^2\int_r^R{\rm dr'}\,r'
\end{equation}
and then
\begin{equation}
p(r)=p_{\rm ext}+{2\pi\over3}G\varrho^2R^2\left(1-\Bigl({r\over R}\Bigr)^2\right)
\,.
\end{equation}
For the central pressure $p_{0}$ at $r=0$ we have numerically
\begin{equation}
p_{0}=p_{\rm ext}+1.258\times10^{-2}\,\left(\varrho\over 3\rm g\,cm^{-3}\right)^2
\left(R\over1\,\rm km\right)^2\ [{\rm bar}]\,.
\label{GasPressInPlan}
\end{equation}
The external pressure in the accretion disk is shown in Fig.~\ref{FigTempDisk}b
and we see that already for bodies as small as 1\,km radius the central pressure
in the body is significantly higher than the pressure in the accretion disk. In
bodies with radii of 10\,km and bigger one encounters central pressures of the
order of 1\,bar and higher.

\begin{figure}

\centerline{
\includegraphics[width=.9\hsize]{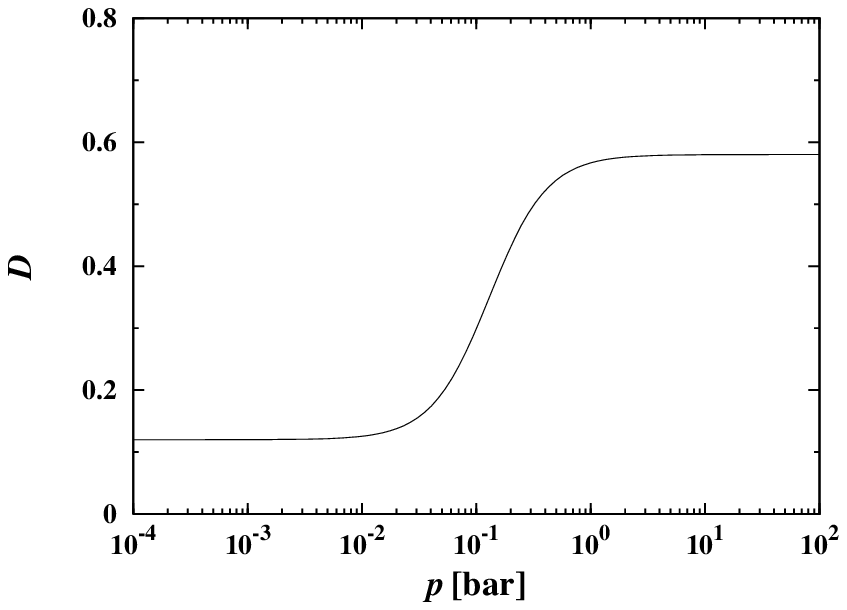}
}

\centerline{
\includegraphics[width=.9\hsize]{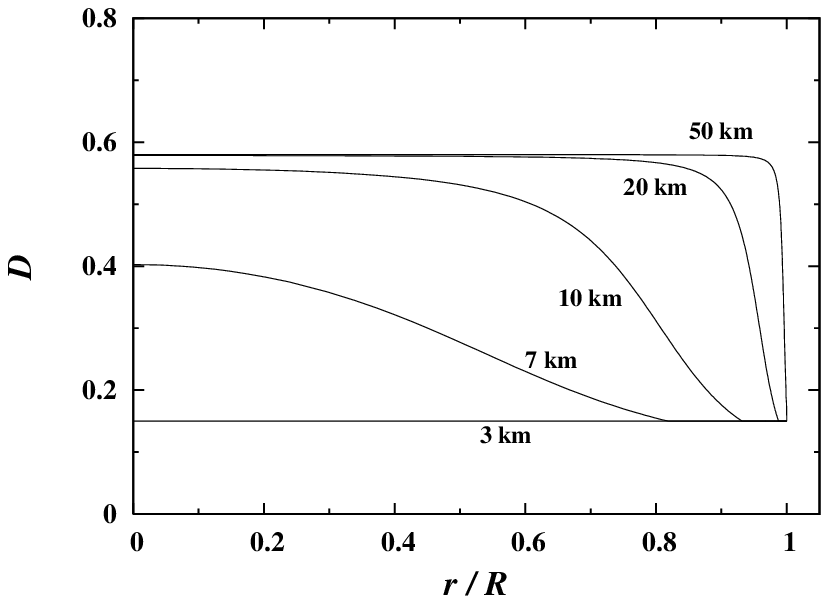}
}

\caption{Relation between relative density (filling factor) $D$ and maximum
experienced pressure for isostatic pressing (top) and run of relative density
within planetesimals of the indicated size (bottom) that results from
cold pressing due to self-gravity.}

\label{FigPhiPIso}
\end{figure}

\subsection{Isostatic pressing of the granular material}
\label{SectColdPress}

The behaviour of granular material under pressure may be rather complex. This
has been discussed by \citet{Gue09} with particular emphasis on impact
processes during planetesimal growth. They performed also laboratory experiments
for the behaviour of fine grained material under the action of static pressure
and how porosity is reduced if the material is compacted by pressing. They
derived a relation  between the applied pressure $p$ and the relative density 
(filling factor) $D$ that is observed after the material has come to rest%
\footnote{Note that they denote the filling factor as $\phi$}
\begin{equation}
D(p)=0.58-0.46\left[\left(p\over p_{\rm m}\right)^{1.72}+1\right]^{-1}\,,\ 
p_{\rm m}=0.13\,{\rm bar}\,.
\label{PorosStatPress}
\end{equation}
This relation only holds for $D>0.15$ because $D\approx0.15$ was the initial
value of the relative density of the material in the experiments before
pressing. Its validity is also limited to pressures where the granular material
is not yet compacted to the relative density $D\approx0.64$ of a random densest
packing of equal sized spheres.

The meaning of Eq. (\ref{PorosStatPress}) is that it describes the porosity
$\phi=1-D$ of a fine powder that experienced the maximum pressure $p$. With
respect to the material in a planetesimal it gives the porosity distribution
within the planetesimal as function of depth for those parts where Eq.
(\ref{PorosStatPress}) predicts a lower porosity than the porosity of the
surface material. The latter is determined by the continued impact processes
during particle growth that result also in a compaction of the powder material
\citep{Wei09}, a process, however, that is not described by 
Eq.~(\ref{PorosStatPress}). The
porosity $\phi_{\rm srf}$ of the surface layer material has to be described in
a different way. Hence we have to determine the porosity in the planetesimal
from the following equation
\begin{equation}
\phi=\cases{
1-D(p) & if\quad $D(p)>1-\phi_{\rm srf}$\cr
\phi_{\rm srf} & else
}.
\label{RelPhiPIso}
\end{equation}
This holds before sintering (see Sect. \ref{SectSinter}) becomes active.
Figure~\ref{FigPhiPIso} shows in the upper part the variation of relative
density $D$ with maximum experienced pressure according to Eq. 
(\ref{RelPhiPIso}). The lower part shows as example the distribution of
relative density within planetesimals with 3\,km to 50\,km radius resulting from
cold isostatic pressing, if no upper limit for the surface porosity is
prescribed. The solutions are obtained by integration of Eq. (\ref{EqHydro})
with Eqs. (\ref{EqDensD}) and (\ref{PorosStatPress}) as equation of state.
This shows that up to planetesimal sizes of about 7\,km the material is not
substantially compacted by self-gravity of the body. For 
planetesimals bigger than 10\,km most part of the body is already compacted by
cold isostatic pressing due to self-gravity to about a density close to that of
the densest random packing of equal sized spheres. Only a rather shallow surface
layer of highly porous material remains in such bodies. This is anyhow evident,
but here a quantitative prediction can be made.

\subsection{Exosphere}

During the first few million years all bodies in a nascent planetary system are
immersed in the protoplanetary accretion disk. Sufficiently massive bodies are
able to gravitationally bind some of the gas. The minimum requirement for this
is that the escape velocity of a gas particle from the surface of a body exceeds
its average kinetic energy of thermal motion corresponding to the disk 
temperature
\begin{equation}
{2GM_R\over R}>{kT\over m_{\rm g}}\,,
\end{equation}
where $R$ is the planetesimal radius and $m_{\rm g}$ the average mass of the gas
particles.
From this one has for the lower limit of the planetesimal radius from which on
gravitational bonding of an atmosphere starts to become possible
\begin{equation}
R_{\rm atm}={2GM_Rm_{\rm g}\over kT}\,.
\end{equation}
For a body with constant density $\varrho$ this means that the radius has to
exceed a radius of
\begin{equation}
R_{\rm atm}=\sqrt{3kT\over 8\pi G\varrho m_{\rm g}}=650\,\left({T\over300\,{\rm K}}\,
{3{\rm g\,cm^{-3}}\over\varrho}\right)^{1\over2}\, {\rm km}
\end{equation}
in order to gravitationally bind gas from the accretion disc and increase the
gas pressure at their surface over the local pressure in the disk. This is only
relevant for protoplanets with radii $R\gtrsim1\,000$\,km. 

\section{%
Temperature structure of planetesimals}

\subsection{%
Heat conduction equation}

For the kind of bodies that we will consider, the material has the structure of
granular matter. The internal structure of such kind of material is not
isotropic and its properties are subject to strong local variations on the
scale of particle sizes. As a result the temperature also will show local
variations on the same length scales. The particles that form the granular
material, however, are very small and in particular they are extremely small
compared to typical length scales over which macroscopic properties of the
bodies may vary. In this case we may average the microscopic properties of the
material and also the temperature over volumes that contain a big number of
particles and at the same time have dimensions small to the characteristic
scale lengths for changes of the values of variables like average temperature
$T$ or average density $\varrho$. Then we work only with such average
quantities, for which we can assume that they are isotropic after averaging
over all possible particle orientations.

With this approximation the equation of energy conservation for a spherically
symmetric body, expressed  as an equation for the temperature $T$ of the
matter, averaged over the microscopic fluctuations, is
\begin{equation}
\varrho c_v\oder Tr+{1\over r^2}\pd{}r\,r^2q_r=+\varrho h
-P\oder{}t\,{1\over\varrho}+\varrho v_r F_r\,,
\label{EqT0}
\end{equation}
where $c_v$ is the average specific heat of the granular material per unit
mass, $q_r$ is the radial component of the average heat flow vector,
$h$ is the average source term for heat production or consumption per unit
mass, the $p$d$V$-term is the compressional work done per unit mass, and the
last term is the work done by external forces. It is assumed that the external
forces,  $F_r$, are species independent and that no differential motion between
the components of the material occurs (flow of gas or water through the porous
matrix is presently not considered). The last two terms have not been
considered so far in model calculations for thermal evolution of asteroids
since they are negligible if the material is practically incompressible.
However, if sintering is considered, the material becomes strongly compressed
during the course of evolution and these terms have to be included.

The quantity $v_r$ is the uniform radial velocity of the mixture components. If
shrinking of the body by sintering is the only kind of motion of the otherwise
stationary structure of the body, the velocity $v_r$ is obtained by
differentiation of Eq.~(\ref{DefMr}) for fixed $M_r$ with respect to time.
There follows
\begin{equation}
v_r=-{1\over r^2}\int_0^r{\rm d}r'\,r'^2\pd{\varrho}{t}\,.
\end{equation}

We will consider models of planetesimals that are in hydrostatic equilibrium
without internal flows. There may be some very slow radial motion of the matter
if the material starts to shrink by sintering at high temperatures. This kind 
of extremely slow motion is completely negligible in the substantial
derivatives d/d$t=\partial/\partial t+v_r\partial/\partial r$ in 
Eq.~(\ref{EqT0}). However, one consequence of shrinking is not negligible: If
the body shrinks the gravitational potential energy decreases and the
corresponding amount of energy is transferred to the matter as heat. This
is described by the term $\varrho v_r F_r$ where $F_r$ is the local 
gravitational acceleration
\begin{equation}
g=-{GM_r\over r^2}\,.
\end{equation}
The term corresponding to the work done by the forces has to be retained. With
this approximation we have the following equation for the temperature
\begin{equation}
\varrho c_v\pd Tr+{1\over r^2}\pd{}r\,r^2q_r=+\varrho h
-P\pd{}t\,{1\over\varrho}-\varrho v_r {GM_r\over r^2}\,,
\label{EqT}
\end{equation}
The variation of $\varrho$ with time is discussed in Sect. \ref{SectSinter}.
 
The heat flow vector has contributions from a number of processes. For the
solid component of the material there is a contribution from heat conduction by
phonons or, in the case of electric conductors (e.g. iron), from conduction
electrons. For a porous material there is also a contribution from the heat
conduction by the gas in the pores. If the material is translucent then one has
also to consider a contribution to heat conduction by radiative transfer. All
these processes have the property, that their contribution to the total heat
flow is proportional to the gradient of the temperature. Generally the
coefficient of proportionality is a second rank tensor, except if the
properties of the material are isotropic, in which case it degenerates to a
simple scalar factor. For granular material the local transport properties for
heat are by no means isotropic. We assume, however, that after averaging the
average heat flow vector is proportional to the gradient of the average
temperature. The radial component of the heat flow vector then takes the
specific form
\begin{equation}
q_r=-K\pd Tr
\end{equation}
with some average heat conduction coefficient $K$ that is different for each
of the different transport processes.

The essential material properties that enter into equation (\ref{EqT}) are
the specific heat per unit mass, $c_v$, the heat conductivity $K$, and the
heat production. In the following subsections we describe how we determine
these quantities for the material of the parent bodies of ordinary chondrites.

The body is also subject to heat and matter exchange with its environment.
This is treated by defining appropriate boundary conditions for Eq.~(\ref{EqT})
that are discussed after our discussion of the material properties.

\begin{figure}

\centerline{
\includegraphics[width=\hsize]{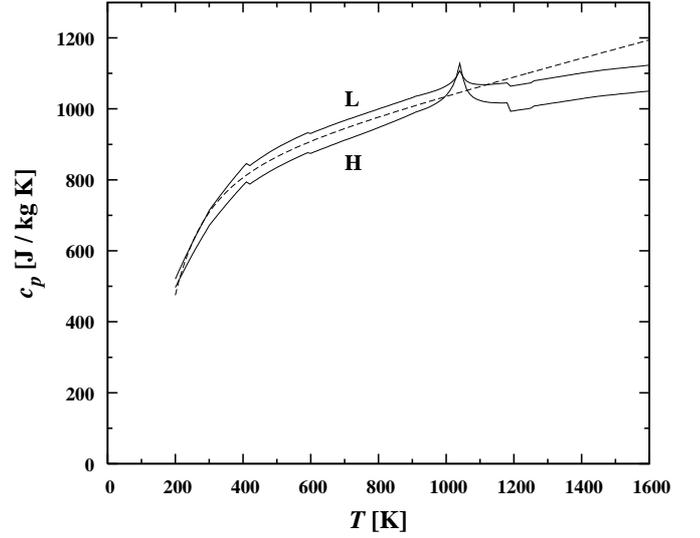}
}

\caption{Specific heat of planetesimal material for mineral mixtures of H chondrites
and L chondrites. Dotted line is the analytic approximation of \citet{Yom84}.}
\label{FiCpMixture}
\end{figure}

\subsection{%
Heat capacity of material}

The heat capacity $c_v$ of a mineral mixture is simply the weighted sum of the
heat capacities of its components. It is calculated in our model calculations
from
\begin{equation}
c_v=\sum_iX_{{\rm min},i}c_{v,i}\,,
\end{equation}
where $X_{{\rm min},i}$ is the mass-fraction of the $i$-th component in the
mixture of solid components and $c_{v,i}$ the heat capacity per unit mass of
component $i$. The quantities $X_{{\rm min},i}$ are given in 
Table~\ref{TabMinMix}. Since we intend to consider bodies of a size of no more
than a few 100\,km, pressures remain below the kbar-range (see 
Eq.~\ref{GasPressInPlan}) and compression of solid material under pressure
is negligible because of the low compressibility of minerals. Under these
conditions there is no significant difference between $c_p$ and $c_v$ and we
may use $c_p$ instead of $c_v$. For our model calculations data for $c_{p,i}(T)$
are taken from the compilation of \citet{Bar95} that gives the heat capacities
$c_p$ per mole. These quantities are converted to heat capacity per unit mass
by dividing by the mole mass $M_i$ of species $i$:
\begin{equation}
c_v=\sum_iX_{{\rm min},i}\,{c_{p,i}\over M_i}\,.
\label{EqHeatCap}
\end{equation}
Values of $c_v$ for the required temperatures are determined by interpolation
in the tables for $c_{p,i}$. The heat capacity for solid solutions is calculated
as weighted mean of the heat capacities of the pure components taking their
mole fractions as weights. 

The variation of the specific heat of the mixture is shown in 
Fig.~\ref{FiCpMixture}. Since some of the minerals suffer structural 
transitions at certain temperatures and since $c_p$ of iron has a cusp at the
Curie-temperature of 1042\,K, the temperature variation of $c_p$ shows some
kinks and jumps \citep[cf. also][]{Gho99}. They might be sources of numerical
problems. For comparison the figure also shows the analytic approximation for
$c_p$ for the bulk material given by \citet{Yom84}. This is an alternative if
the jumps in the temperature variation of $c_p$ result in  numerical problems,
but in our calculations it was not necessary to take recourse to this
approximation. 

\begin{table}
\caption{Data for calculating heating rates by decay of radioactive nuclei.}

\centerline{
\begin{tabular}{l@{\hspace{.5cm}}ll@{\hspace{.1cm}}lll}
\hline\hline
\noalign{\smallskip}
Species & \multicolumn{1}{c}{$f$} &  \multicolumn{1}{c}{$E$} & 
& \multicolumn{1}{c}{$\tau$} & \multicolumn{1}{c}{$h$} \\
   &   &  [MeV] & & \multicolumn{1}{c}{[a]} & \multicolumn{1}{c}{W\,kg$^{-1}$} \\
\noalign{\smallskip}
\hline
\noalign{\smallskip}
$^{26\!}$Al   & $5.1\times10^{-5}$ & 3.188 & (1) & $1.0\times10^6$ & 
    $1.67\times10^{-7}$\\
$^{60}$Fe (high) & $1.6\times10^{-6}$ &   2.894 & (1) & $3.8\times10^6$ &
    $2.74\times10^{-8}$ \\
\phantom{$^{60}$Fe} (low)           & $4.2\times10^{-7}$ &         &                 &  \\
$^{40}$K    & $1.5\times10^{-3}$ &  0.693 & (2) & $1.8\times10^9$ &
    $2.26\times10^{-11}$ \\
$^{232}$Th  &               1.00 &  40.4  & (2) & $2.0\times10^{10}$ &
    $1.30\times10^{-12}$ \\
$^{235}$U   &               0.24 &  44.4  & (2) & $1.0\times10^9$ &
    $3.66\times10^{-12}$ \\
$^{238}$U   &               0.76 &  47.5  & (2) & $6.5\times10^9$ &
    $1.92\times10^{-12}$ \\
\noalign{\smallskip}   
\hline
\end{tabular}
}

\bigskip
{\scriptsize
Sources: (1) \citet{Fin97}, (2) \citet{VSc95}
}

\label{TabHeatDec}
\end{table}

\subsection{%
Heating by radioactive nuclei}
\label{SectHeatRate}

Next we consider the source term $h$ in Eq.~(\ref{EqT}). There are essentially
two different kinds of sources and sinks of heat within the planetesimal
bodies. One source is the energy liberated during decay of radioactive isotopes
of a number of elements. The other one is consumption of energy during melting
of planetesimal material or liberation of energy during solidification of the
melt. Melting is not considered because we aim to study parent bodies of
undifferentiated meteorites.

The main sources of heat input by radioactives during the early heating up of
planetesimals and the subsequent cooling phase are decay of $^{26\!}$Al and
possibly $^{60}$Fe \citep[cf. the discussion in][]{Gho06}. More long-lived
radioactive nuclei,  essentially $^{232}$Th, $^{235,238}$U, and $^{40}$K, are
responsible for the later long-term evolution of the temperature. For the
nuclei decaying by $\beta$-decay ($^{26\!}$Al, $^{60}$Fe, $^{40}$K) the energy
of the fast electrons and of the emitted $\gamma$-photons is absorbed within
the planetesimal material and is converted to heat, while the neutrinos leave
the bodies and carry away their part of the energy. For the nuclei decaying by
$\alpha$-decay the whole decay energy is absorbed by the planetesimal material.
We assume that no chemical differentiation occurs in the bodies that we
consider. Hence, after having averaged over the inhomogeneous microstructure
of the material, the heat producing nuclei are homogeneously distributed over
the bodies. The heat production rate by these nuclear processes is
\begin{equation}
h_{\rm nuc}=\sum_i {X_{{\rm el},i}\over m_{{\rm el},i}}f_i{E_i\over\tau_i}\,
{\rm e}^{-t/\tau_i}\,.
\end{equation}
The sum is over all nuclei that contribute to heating, $X_{{\rm el},i}$ denotes
the mass-fraction of the corresponding element in the material of the
planetesimals (see Table~\ref{TabMinMix}), $m_{{\rm el},i}$ the atomic mass of
the element for the isotopic mixture at the time of formation of the solar
system, $f_i$ the fraction of the isotope of interest at the time of formation
of the solar system, $\tau_i$ the decay time scale for e-fold decrease of
abundance of the isotope, and $t$ is the time elapsed since formation of the
solar system.
   
\begin{figure}

\includegraphics[width=\hsize]{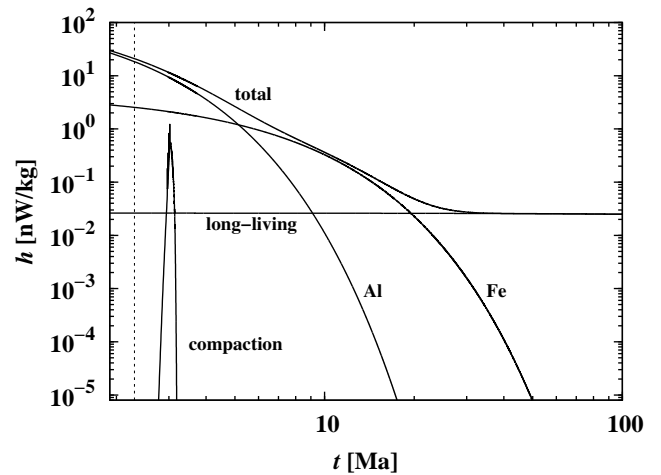}

\caption{Contributions of different heating mechanisms to the total heating
rate. Time $t$ is after formation of CAIs. The dashed line indicates the
formation time of the H chondrite parent body (2.3\,Ma, see 
Sect.~\ref{SectFitHChond}). The release of gravitational energy by contraction
is shown for the final model of the H chondrite parent body.
}

\label{FigHeatRate}
\end{figure}

The constants for calculating $h_{\rm nuc}$ are given in Table~\ref{TabHeatDec}.
The element abundances used for calculating $X_{{\rm el},i}$ for the
mineral mixture given in Table~\ref{TabMinMix} are taken from \citet{Lod09}.
Isotopic abundances of K and U at time of solar system formation are taken from
\citet{And89}. The abundance of $^{26\!}$Al is that given by \citet{Nyq09}.
The abundance of $^{60}$Fe is disputed in the literature. Table \ref{TabHeatDec}
gives the probably uppermost value from \citet{Bir88} and the lower limit
according to \citet{Qui07}. There are indications that $^{60}$Fe was not
homogeneously  distributed in the early solar system \citep{Qui10}, which means
that the initial $^{60}$Fe abundance in the parent bodies is not known a priori
and is an additional free parameter for the modelling. For the decay time of
$^{60}$Fe the recent revised value for the half-lfe $\tau_{1/2}=2.62\pm0.04$\,Ma of
\citet{Rug09} is used.

Figure \ref{FigHeatRate} shows the variation of the heating rate per unit
mass with time elapsed since formation of CAI, calculated with the high $^ {60}$Fe
abundance (see Table~\ref{TabHeatDec} for this). The domi\-nant heating source
is $^{26}$Al at the time of formation of planetesimals ($t\lesssim5$\,Ma),
but $^{60}$Fe dominates as heat source for an extended period from 
$\approx5$\,Ma to $\approx20$\,Ma because of the revised $^{60}$Fe half-life.

For comparison Fig.~\ref{FigHeatRate} also shows the contribution of the release
of gravitational energy to heating during shrinking of the body, resulting for
the model of Sect.~\ref{SectFitHChond}. For the rather small bodies that are
considered in this paper this heating source is not important (but included
in the model).

\subsection{%
Heat conduction by the porous solid material}
\label{SectApptHeatK}

For the heat conductivity $K$ of the chondritic material we use two different
types of experimental data. For low porosities from the range $0<\phi\le0.2$ we
use data measured for a number of ordinary H and L chondrites by \citet{Yom83}.
For high porosities $\phi>0.4$ we  use the data for silica powder derived from
laboratory measurements by \citet{Kra11}. All these measurements were conducted
under vacuum conditions in order to exclude any contribution from heat
transport from gas-fillings in the pores. 

\begin{figure}

\includegraphics[width=\hsize]{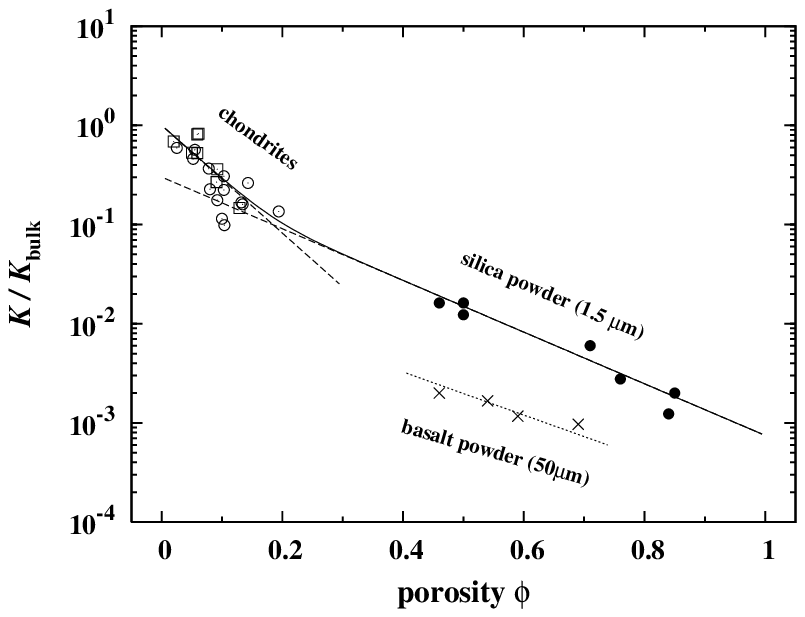}

\caption{Variation of heat conductivity K with porosity $\phi$. Results
for fine grained silica powder (filled circles) from experiments of 
\citet{Kra11,Kra11b}, and for particulate basalt (crosses) from \citet{Fou70}.
Typical grain size is indicated for both cases. Open squares and open circles
are experimental results for heat conductivity and porosity for ordinary H and L
chondrites, respectively,  from \citep{Yom83}. Solid line is analytic fit, 
Eq.~(\ref{FitToKpor}), to the data.}

\label{FigKpor}
\end{figure}

Figure \ref{FigKpor} shows conductivity $K$ plotted versus porosity for chondritic
material at $T=300$\,K. The data for H and L chondrites scatter significantly
and because of the small number of available data points no obvious systematic
difference between the two types of material can be recognized. Therefore we fit
both sets of data with a single analytic approximation
\begin{equation}
K_1(\phi)=K_{\rm b}{\rm e}^{-\phi/\phi_1}
\label{Fit1Kphi}
\end{equation}
with two constants $K_{\rm b}$ and $\phi_1$. This exponential form enables a
reasonable fit of the data. The constant  $K_{\rm b}$ may be interpreted as the
extrapolated average thermal conductivity of the bulk material at vanishing
porosity, for which we obtain $K_{\rm b}=4.3$\,W\,m$^{-1}$\,K$^{-1}$.  For the 
second constant we choose $\phi_1=0.08$ \citep[see also][]{Kra11b}. In Fig. 
(\ref{FigKpor}) the data are normalized with the value of $K_{\rm b}$. The
dashed line running through the data points for the chondrites give our
approximation $K_1(\phi)$.

At high porosities Fig. (\ref{FigKpor}) shows the conductivity $K$ for a silica
powder that consists of equal sized spheres with $1.5\,\mu$m diameter. By the
nature of the experimental method the data of \citet{Kra11} do not refer to a
well defined temperature but the heat conductivity was derived by analysing the
cooling behaviour of their sample. Therefore the value of $K$ is some average value
over the raise and fall of the temperature in the experiments, which is somewhat
above room temperature. The data are fitted with an analytic approximation of
the form
\begin{equation}
K_2(\phi)=K_{\rm b}{\rm e}^{a-\phi/\phi_2}
\label{Fit2Kphi}
\end{equation}
with two constants $a$ and $\phi_1$ and the same value of $K_{\rm b}$ as before. 
This type of exponential dependence on $\phi$ allows a reasonable fit of the
data points also in this case. The constants are found to be $a=1.2$ and 
$\phi_2=0.167$ \citep[in an earlier version,][slightly different values of the
constants were given]{Kra11b}. This fit is shown as the second dashed line in 
Fig.~\ref{FigKpor}. 

The experiments of \citet{Kra11} are conducted with very fine grained silica
powder. This is not the same kind of material as it is found in chondrites, but
for two reasons it may be considered as a reasonable proxy for chondrite
material before strong compaction. First the mineral mix in chondrites is
dominated by silicates and all silicates have similar heat conduction
coefficients. Second the heat conduction of very loosely packed material is via
the tiny contact regions between the particles. In chondrites part of the
material are the rather big chondrules (size $\approx1$\,mm), but as long as
the material is not strongly compacted and the chondrules are well separated
by the very small grained matrix (particle sizes $\lesssim0.5\,\mu$m), the
heat conduction is obviously governed by the heat flow through the contact
points between the tiny matrix particles. In this respect the basic physics
of the heat transport in the experiment of \citet{Kra11} should be very similar
to that in chondrite material before strong compaction.

The powder particles used in the experiments are about a factor of five to ten
times bigger than the matrix particles in chondrites \citep{Rie93}. Some 
indications on the influence of particle sizes may be obtained by comparing
the results of \citet{Kra11} with
results of heat conduction measurements of \citet{Fou70} for powders of 
basaltic particulates that are much coarser grained. Figure~\ref{FigKpor} shows
results for their size separate with average particle size of $50\,\mu$m. The
variation of $K$ with porosity for the \citet{Fou70} granular material is
very similar as of the silica powder used by \citet{Kra11}, except that the
conductivity is lower by a factor of somewhat less then a factor of ten. The
particle sizes, on the other hand, are bigger on average by more than a factor
of thirty. This suggests that the heat conductivity measured by \citet{Kra11}
for the highly porous silica powder underestimates the conductivity of the
matrix material of chondrites at the same porosity, but probably not by a big
factor. In case of a power law dependence of $K$ on particle sizes one may 
speculate that a by a factor of about three higher conductivity of meteoritic
material than for the silica powder may be an appropriate estimation, but
without more definite information we take the values of \citet{Kra11} for our
model calculations.  
  
\begin{figure}

\includegraphics[width=\hsize]{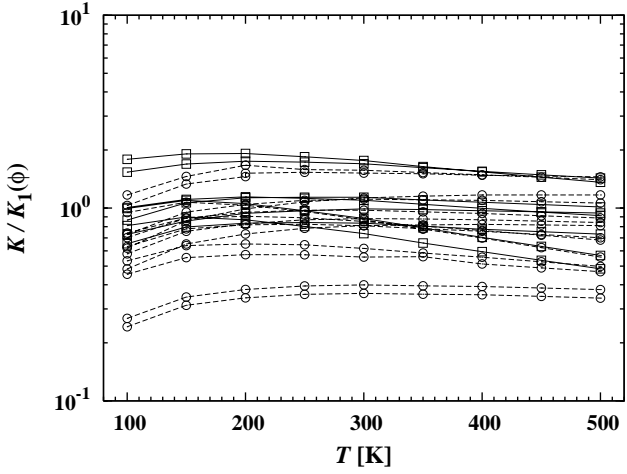}

\caption{Temperature variation of thermal conductivity $K$. Plotted are data
for H chondrites (open rectangles) and L chondrites (open circles), normalized
with the $\phi$-variation according to approximation (\ref{Fit1Kphi}).  The
lines connect data for each of the meteorites.
}
\label{FigTvarKb}
\end{figure}

The two fits, Eq. (\ref{Fit1Kphi}) and Eq. (\ref{Fit2Kphi}), for $\phi<0.2$ and
$\phi>0.4$, respectively, are combined into a single analytic approximation 
for $K(\phi)$ by
\begin{equation}
K(\phi)=\left(K_1^4(\phi)+K_2^4(\phi)\right)^{1/4}
\label{FitToKpor}
\end{equation}
in order to smoothly interpolate between the two limit cases, in particular in
the intermediate range of porosities $0.2\le\phi\le0.4$. This approximation is
shown as the full line in Fig.~\ref{FigKpor}.

The bulk conductivity $K_{\rm b}$ in Eq. (\ref{Fit1Kphi}) is temperature
dependent. Figure \ref{FigTvarKb} shows data for $K$ for H and L chondrites as
given by \citet{Yom83}, divided by $K_1(\phi)$, given by Eq.~(\ref{Fit1Kphi}).
The data for $K(T)/K_1(\phi)$ show for each of the meteorites a clear systematic 
variation with temperature. The extent of these variations, however, is much
less than the scattering between the different meteorites which amounts to 
variations by a factor of about two. The origin of these meteorite-to-meteorite
variations is not known, but most likely they have their origin in
variations in the composition and structure of the meteoritic material. Such
variations can presently not accurately be accounted for and will not be
considered in our model calculations. Therefore we will also neglect the small
variation of $K$ with $T$ and take $K_{\rm b}$ in our calculations to be 
temperature independent.

\subsection{%
Heat conduction by other processes}

The material of the planetesimals is dominated by minerals that are transparent
in the far infrared region. Therefore energy transport by radiation is possible.
The energy flux by radiation increases strong with increasing temperature and
for temperatures of the order of about 800\,K and higher the heat flux by
radiation becomes non-negligible. The complicated structure of the material
(chondrules densely packed in a porous dust matrix) makes it difficult to
calculate this from first principles. We follow therefore the proposal of 
\citet{Yom84} to take the results of the laboratory measurements of 
\citet{Fou70} of radiative heat conduction in basaltic powders as an
approximation for meteoritic material. 
The model calculations show that for the bodies of interest (temperature
stays below melting temperature) radiative heat conduction never becomes an
important energy transport mode. This is because if temperature becomes high
enough for radiative heat conduction to contribute somewhat to the net heat
flux, the onset of sintering above $\approx700$\,K results in a strong
increase of heat conductivity by phonons that outnumbers radiative contributions
again.

A possible contribution to heat conduction by pore gas is small and neglected in
this paper; details of this will be discussed elsewhere.
  
\subsection{%
Boundary conditions}

For solving the heat conduction equation (\ref{EqT}) one has to specify an
initial condition $T(r,t_0)$ at some initial instant $t_0$ and boundary
conditions, in our case at $r=0$ and $r=R$. 

The inner boundary condition is that there is no point source for heat at the
centre which translates into the Neumann boundary condition
\begin{equation}
\left.\pd Tr\right\vert_{r=0}=0\,.
\label{HeatBoundCentre}
\end{equation}

The temperature $T_{\rm s}$ of the surface layer of the body is determined by
the equilibrium between (i) the energy fluxes  toward the surface from the
interior and from the outside, and (ii) the energy fluxes away from the surface
to exterior space \citep[cf., e.g.,][]{Gri89,Gho03}. There results a mixed 
boundary condition at the surface
\begin{equation}
-K\left.\pd{T}r\right\vert_{r=R}=-\sigma_{\rm SB} T_{\rm s}^4
+F_{\rm ext}\,.
\label{BoundCondTsNeum}
\end{equation}
The left hand side describes the energy flow from the interior toward the
surface by heat conduction.
The first term on the r.h.s. is the rate of energy loss by radiation from the
surface to exterior space, the second term, $F_{\rm ext}$, is the rate of energy
gain by outer sources. During the first few million years, if the planetesimals
are still embedded in an optically thick accretion disc, $F_{\rm ext}$ is
given by $\sigma_{\rm SB} T_{\rm c}^4$, with $T_{\rm c}$ being the temperature at
the midplane of the disk (see Fig.~\ref{FigTempDisk}a). After disk dispersal the
planetesimal is irradiated by the proto-sun and the rate of energy input, 
$F_{\rm ext}$, is given by $(1-A_{\rm surf})\sigma_{\rm SB} T_*^4R_*^2/4a^2$.
Here $a$ is the (average) distance to the star and $A_{\rm surf}$ is the albedo
of the planetesimals surface.

Alternatively one can consider at the surface the Dirichlet condition 
\begin{equation}
T(R)=T_{\rm b}
\label{HeatBoundSurfT}
\end{equation}
with some prescribed value $T_{\rm b}$. This has been done in a number of
published model calculations where some fixed value for $T_{\rm b}$ was
assumed.

\subsection{%
Initial condition}

Large planetesimals with radii of the order of 100\,km occur as transition
states during the growth from km sized bodies of the initial planetesimal swarm
to protoplanets with sizes of the order of 1\,000\,km. The growth initially
proceeds rather slowly on timescales of a few 10$^5$ years until run-away
growth commences after the biggest planetesimals reach sizes of about 
10-20\,km \citep[e.g.][]{Wei06,Naga07}. During run-away growth the mass rapidly
increases within less than 10$^5$ years to the size of a protoplanet. This
means that the bodies are formed on timescales shorter than the decay time of
$\approx1$\,Ma for $^{26\!}$Al and also that they collect most of their mass
within a period even much shorter than this. There is not sufficient time
available for strong heating by $^{26\!}$Al decay of those planetesimals that
contributed to the growth of a 100\,km sized body; most of the heat released by
radioactive decay is released after its formation. 

Therefore we will base our calculations in this paper on the ``instantaneous
formation'' approximation where it is assumed that the body is formed within
such a short period that all heating occurred after its formation.
Within the framework of the instantaneous formation approximation it is
appropriate to prescribe the temperature $T_{\rm c}$ of the disk material at
the formation time of the planetesimals as initial value of the temperature.

Modifications resulting from a finite duration of the growth period will be
considered elsewhere.

\section{%
Sintering}
\label{SectSinter}

The compaction of material in planetesimals is a two-step process.  The 
initially very loosely packed dust material in the planetesimals comes under
increasing pressure by the growing self-gravity of the bodies. The granular
material can adjust by mutual gliding and rolling of the granular components to
the exerted force and evolves into configurations with closer packing. The
ongoing collisions with other bodies during the growth process enhances this
kind of compaction of the material. This mode of compaction, ``cold pressing'',
by its very nature does not depend on temperature and operates already at low
temperature; it was considered in Sect.~\ref{SectColdPress}. A second mode of
compaction commences if radioactive decays heat the planetesimal material to
such an extent, that creep processes are thermally activated in the lattice of
the solid material. The granular components then are plastically deformed under
pressure and voids are gradually closed. This kind of compaction by ``hot
pressing'' or ''sintering'' is what obviously operated in ordinary chondrite
material and the different petrologic types 4 to 6 of chondrites are obviously
different stages of compaction by hot pressing.

\citet{Yom84} were the first to perform a quantitative study of this process
by applying early theories of sintering developed in material science.
We follow here essentially the same approach because more advanced modern
theories of hot pressing are developed to model metallurgical processes that
apply generally much higher pressures ($\gg1$\,kbar) than what is typically
encountered in compaction of planetesimal material ($\lesssim10^2$\,bar) and
are mainly concerned with the final stages of the process. Because the rate of
increase of temperature in planetesimals is very low (of the order of 
$10^{-3}$\,K\,yr$^{-1}$) the creep processes result in finite deformations of 
the material already at rather low temperatures or pressures, where under
laboratory conditions no effect is seen. The more simple early theories of hot
 pressing seem better to fit to such situations.

\subsection{Equations for hot pressing}

For describing the sintering process we initially assume a dense packing of
equal sized spheres with initial radius $R_0$. The packing is sufficiently
dense that no further compaction can be achieved by pressing without crushing
the spheres. On average, the individual granular units will touch each other
at $Z$ contact points. At sufficiently high pressure and temperature the
individual spheres will plastically deform at the contact points by creep
processes and contact faces will form between adjacent particles while the
volume of the particle will remain constant. As sintering proceeds, the 
voids between the spheres become smaller and the sphere centres get closer.

There are two stages for this process. In the first stage the voids form an
interconnected network between the granular units. This closes at some stage
of the sintering process and there remain isolated pores, that have to close
by further sintering in a second stage. The relative density $D$ at the 
transition between stages one and two depend on the type of packing. The
following approximations are for the first stage.

The first basic assumption of the deformation theory of hot-pressing by 
\citet{Kak67}, on which the work of \citet{Yom84} is based, is a pure 
geometrical one. It is assumed that the formation of the contact faces can be
conceived as if at each contact point a cap would have been cut-off from each
of the two contacting grains. Then, for grains of equal radius, the contact
areas are circular areas with radius $a$. It is assumed that all cut-away caps
have the same height $h$ and radius $a$ at their base. The volume of one such
cap is
\begin{equation}
V_{\rm cap}={\pi\over6}\,h^2\left(3R-h\right)
\label{HotPressMod1}
\end{equation}
and its height 
\begin{equation}
h=R-\sqrt{R^2-a^2}\,.
\label{HotPressMod1a}
\end{equation}
The granular units then are (by assumption) spheres with $Z$ caps cut-off from
them. To conserve the original volume of the sphere, the radius $R$ of such a
truncated sphere has to be bigger than the pristine radius $R_0$. Conservation
of volume requires
\begin{equation}
{4\pi\over3}R_0^3={4\pi\over3}R^3-ZV_{\rm cap}\,.
\label{HotPressVolCons}
\end{equation}
This holds as long as the contact areas do not come into contact with each
other. The relation to the relative density $D$ is \citep{Kak67}
\begin{equation}
\left(\frac{a}{R} \right)^2 = 1- \left(\frac{R_0}{R} \right)^2 
\left( \frac{D_0}{D} \right)^{2/3}.
\label{HotPressDens}
\end{equation} 
Here $D_0$ is the relative density of the initial packing of spheres with
radius $R_0$. For a given number of contact points $Z$ and given $D_0$, $R_0$,
Eqs. (\ref{HotPressMod1}) to (\ref{HotPressDens}) define $R$ and $a$ in terms
of the relative density $D$. 

In the theory of \citet{Rao72} a number of regular packings of spheres is
considered for which the number of contact points $Z$ is fixed ($Z=6,8,12$). In
particular they favour the 'hexagonal prismatic' packing with $Z=8$ and gave
their formula for this case. This is the model that has been used by 
\citet{Yom84} in their study of sintering of planetesimals. They argued that
many experiments on the packing of small spherical particles of constant size
show that the porosity achieved after sufficient tapping would be near 
40\%, with an average of  about eight contact points per particle. Since a
regular hexagonal prismatic packing of spheres also has a coordination 
number of eight and a porosity of 39.5\%,  and a random close packing has
porosity 36\% and on average $Z=7.3$ (see Sect.~\ref{SectGranular}), they used
that packing model in their sintering models. For a discussion of the more general case of random packings see, e.g., \citet{Arz82,Arz83,Fis83}; the
equations obtained in that case are more involved, but there are no basic
differences.

The second basic assumption in the theory of \citet{Rao72} of hot pressing is
that the strain rate is related to the applied stress by the relation 
for power-law creep
\begin{equation}
\dot\epsilon=A\sigma_1^n\,,
\label{DefPowCreep}
\end{equation}
and that $\dot\epsilon$ is given in terms of the rate of change of relative
density as
\begin{equation}
\dot\epsilon=-{\dot D\over D}\,.
\label{RateStrain}
\end{equation}
The stress $\sigma_1$ is the pressure acting at the contact faces of the
granular units. The quantities $A$ and $n$ have to be determined experimentally
for each material. The quantity $A$ depends on temperature.

The stress $\sigma_1$ is given by the pressure acting at the contact areas 
between the granular units. It is assumed that this is given in terms of the
applied pressure $p$ and the areas of contact faces, $\pi a^ 2$, and  average
cross-section of the cell occupied by one granular unit, $C_{\rm av}$, as 
\begin{equation}
\pi a^2\sigma_1=C_{\rm av}p\,.
\label{EffStress}
\end{equation}
In the hexagonal prismatic packing model favoured by \citet{Yom84}, the
cross-section $C_{\rm av}$ is given by
\begin{equation}
C_{\rm av}=2\sqrt{3}\,(R^2-a^2)\,.
\end{equation}
Via the dependence of $R$ and $a$ on $D$ this is a function of $D$. Values for
other packing models \citep[e.g.][]{Kak67} are within O(1) of this. In the 
initial stages of sintering (small $a$) the stress $\sigma_1$ is much higher
than $p$, in the final stages it approaches $p$. 

Equations (\ref{DefPowCreep}), (\ref{RateStrain}), (\ref{EffStress}) result
in the following differential equation for the relative density
\begin{equation}
{\partial\,D\over\partial\,t}=-DA\left({C_{\rm av}\over\pi a^2}p\right)^n\,.
\label{DGLforD}
\end{equation}
With the above relations between $R$, $a$ and $D$ this is a (closed) set
of equations for calculating the time evolution of $D$, that has to be solved
together with the other equations for the structure and evolution of the
planetesimal which define the pressure and temperature.

The transition to stage 2 by closure of voids is assumed in models of hot 
pressing to occur at $D\gtrsim0.9$ \citep[e.g.][]{Arz83}. The equation for
$\dot D$ becomes $\dot D\propto1-D$ for this case \citep[cf.][]{Wil75,Arz83}.
Because the corresponding full equations are similar in structure to 
-Eq.~(\ref{DGLforD}) except for the factor $1-D$ we include the final pore-closing stage
simply by multiplying for $D>0.9$ the r.h.s. of Eq.~(\ref{DGLforD}) by a factor
\begin{equation}
F=10\cdot(1-D) \quad(D>0.9)\,,
\end{equation}
to get a continuous transition between both cases.
 
\subsection{Data for olivine}

The pre-factor $A$ and the power $n$ in Eq.~(\ref{DefPowCreep}) have to be
determined by laboratory exponents. \citet{Yom84} used data from \citet{Sch78}
for olivine. No other data for materials of interest for planetesimals seem to
have been determined since then. \citet{Sch78} gave the following fit to their
experimental data for small spheres of olivine ($R < 53,\mu$m):
\begin{eqnarray}
\dot\epsilon = A \cdot \frac{\sigma^{3/2}}{R^3} e^{E_{\rm act}/
TR_{\rm gas}}, 
\label{S_Sintern}
\end{eqnarray}
with $\sigma$ stress on contact faces (in bar), $E_{\rm act}$ the activation
energy for creep, $T$ the temperature, $R_{\rm gas}$
the gas constant, and $R$ the radius of granular units (in units cm). 

For the activation energy a value of $E_{\rm act}=85\pm29\,\rm kcal\,mol^ {-1}$
was given by \citet{Sch78}. In the model calculations we use the value 
$E_{\rm act}=85\,\rm kcal\,mol^ {-1}$. For the pre-factor $A$ a range of values
from $1.6\times10^{-5}$ to  $5.4\times10^{-5}$ was given by \citet{Sch78}; as a 
compromise we use a value  of $A=4\times10^{-5}$ in our model calculations. 

Note that \cite{Yom84} choose to use for $\dot\epsilon$ the approximation
with $n=1$ given in Eq.~(7) in \citet{Sch78}, while we prefer to use use the
approximation given by Eq.~(8) of \citet{Sch78}, because they explicitly state 
that this describes their measured $\sigma$-dependence.

\section{%
Results for thermal evolution of planetesimals}

\subsection{Model calculation}
\label{SectMethModelCalc}

The calculation of a model requires to solve the differential equations for 
heat conduction, Eq.~(\ref{EqT}), hydrostatic equilibrium, Eq.~(\ref{EqHydro}),
for the evolution of porosity, Eq.~(\ref{S_Sintern}), together with equations
for the material properties, the equation of state,  Eq.~(\ref{EOSPor}), the 
equations for heat conductivity,  Eq.~(\ref{FitToKpor}), and heat capacity,
Eq.~(\ref{EqHeatCap}), and together with appropriate initial and
boundary conditions.

The heat conduction equation and the pressure equation are re-written in terms
of $M_r$, defined by Eq.~(\ref{DefMr}), as independent variable and are
discretised for a set of fixed mass shells with masses $\Delta M_i$ ($i=1,I)$
and shell boundaries $r_i$ ($i=0,I$). The $M_r$-coordinate corresponds to a
Lagrangean coordinate that is fixed to the matter. For this choice of
coordinates there is no flow of matter across cell boundaries. This enables 
a simple treatment of growth of the body, if this is considered, and it
avoids problems with numerical diffusion in case of inhomogeneous composition
(e.g., radial variation of porosity).

\begin{table}

\caption{Model parameters for the model of \citet{Miy81} (MFT81) for a consolidated
body (average L chondrite), and for a simi\-lar model of an initially porous body without
(PL0) and with (PL) additional heating by decay of $^{60}$Fe and long-lived nuclei. 
}

{\small
\begin{tabular}{@{}llllll@{}}
\hline\hline
\noalign{\smallskip}
Quantity &  & MFT81 & PL0 & PL & Units \\
\noalign{\smallskip}
\hline
\noalign{\smallskip}
radius               & $R$               & 85               & 85            & 85    & km \\
formation time       & $t_{\rm form}$    & 2.4              & 2.3           & 2.3   & Ma \\
heat conductivity    & $K_{\rm b}$       & 1                & 4.3           & 4.3  & W\,m$^{-1}$\,K$^ {-1}$ \\
surface temperature  & $T_{\rm s}$       & 180              & 150        & 150      & K \\
density              & $\varrho_{\rm bulk}$ & 3.2              & 3.59         & 3.59   & g\,cm$^ {-3}$  \\ 
$^ {26\!}$Al/\,$^{27\!}$Al &                   & 5 & 5.1 & 5.1 &
$\times10^{-5}$ \\ 
$^ {60}$Fe/$^{56}$Fe &                   &  ---             & 0 & 4.1 &
$\times10^{-7}$ \\
initial porosity     & $\phi_{\rm srf}$  & 0 (10\%)         & 60\%   & 60\%          &  \\
\noalign{\smallskip}
\hline
\end{tabular}
}

\label{TabMiyaParm}
\end{table}

The heat conduction equation is  solved by a fully implicit finite difference
method with Neumann boundary condition, Eq.~(\ref{HeatBoundCentre}), at centre
and Dirichlet boundary condition, Eq.~(\ref{HeatBoundSurfT}), at the surface.
The first order ordinary differential equation for~$\phi$, 
Eq.~(\ref{S_Sintern}), is solved by the fully  implicit Euler method.

\begin{figure}

\centerline{
\includegraphics[width=.8\hsize]{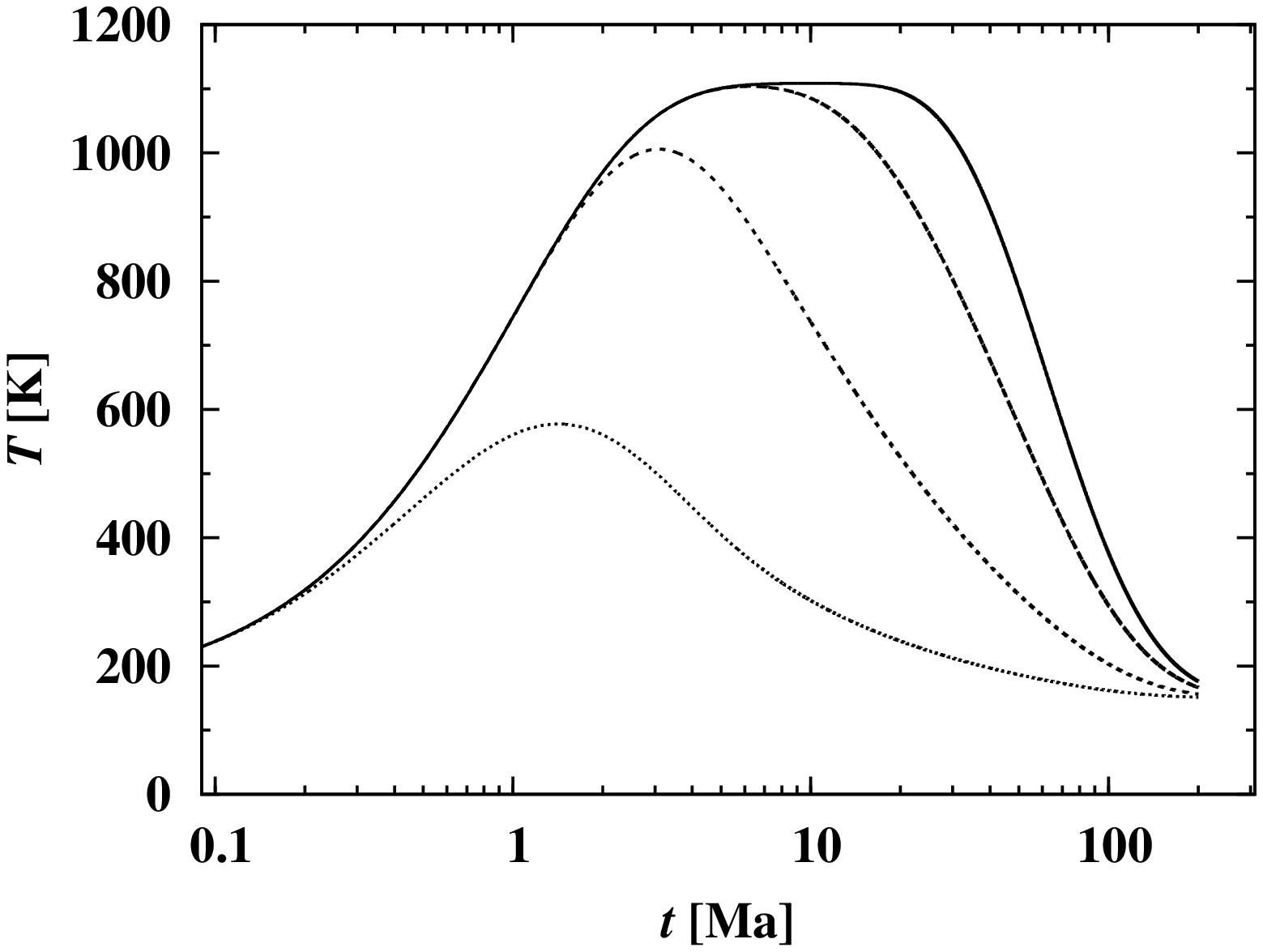}
}

\centerline{
\includegraphics[width=.8\hsize]{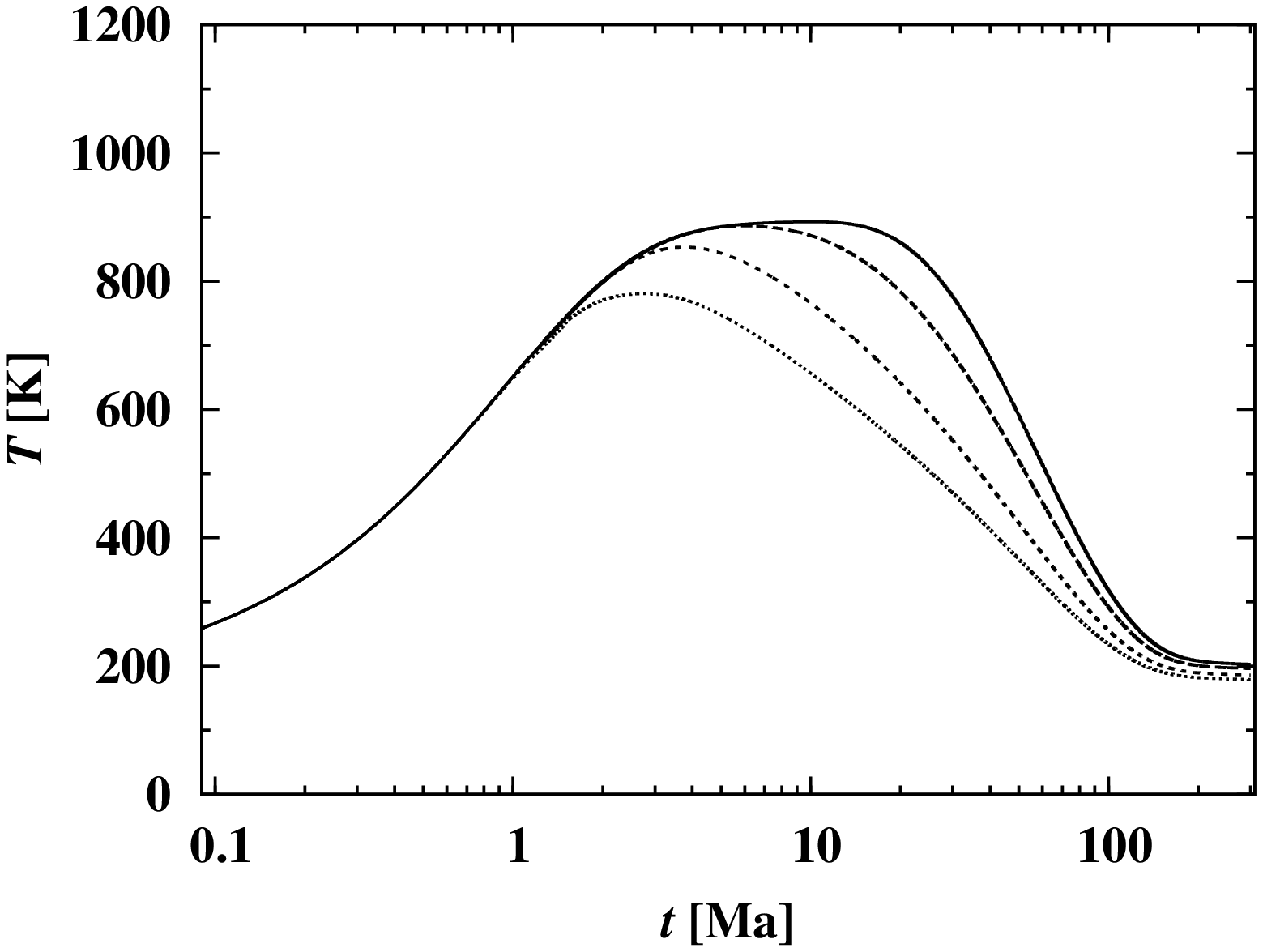}
}

\centerline{
\includegraphics[width=.8\hsize]{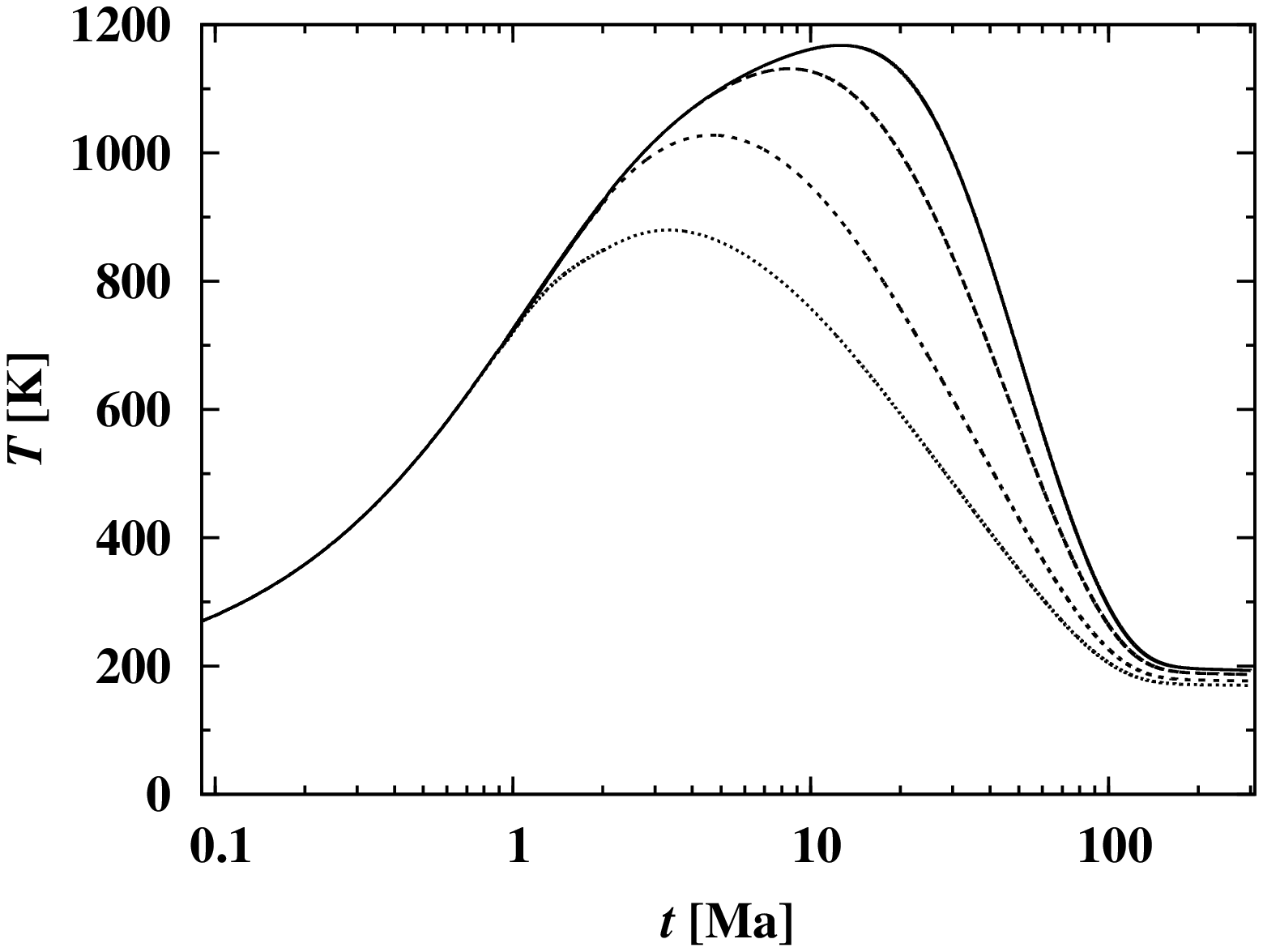}
}

\caption{Temperature evolution of a body of 85\,km radius at different depths
from the surface and at centre: Dotted line at depth 4.25\,km, short dashed line
at 17\,km depth, long dashed line at 23\,km depth. Full line shows temperature
evolution at the centre. {\bf(a)} The model is calculated with the same physical
input as in the analytical model of \citet{Miy81}. Model parameters MFT81
in Table \ref{TabMiyaParm}. {\bf(b)} Similar model, but now calculated for
a porous body, considering thermal conductivity of porous material according 
to Eq.~(\ref{FitToKpor}), and sintering and cold pressing as
described in this paper.
Model parameters PL0 in Table \ref{TabMiyaParm}. {\bf(c)} Same kind of
model as (b), but now additional heating by $^{60}$Fe and long-lived nuclei
considered. Model parameters PL in Table~\ref{TabMiyaParm}.
}

\label{FigTmevolMiya} 

\end{figure}

In order to account for the non-linear coupling between the different equations
we perform a fixed-point iteration where we solve the equations in turn as
follows:
\begin{enumerate}

\item Given are values of $\phi_i$, $T_i$, $\Delta M_i$ for each mass shell $i$
at some instant $t_{k-1}$. We have to calculate new values at next instant 
$t_k=t_{k-1}+\Delta t$. The values of $\phi_i$, $T_i$ at $t_{k-1}$ are used as
starting values for the iterative calculation of $\phi_i$, $T_i$ at $t_k$.

\item The heat production by radioactive decays over the period $\Delta t$ is
calculated for each shell.
 
\item For given porosity $\phi_i$ one finds $\varrho_i$ from Eq.~(\ref{EOSPor})
and with given mass  $\Delta M_i$ in each mass shell $i$ we calculate, starting
from the centre, the shell boundaries $r_i$ at $t_k$.

\item From the change of $r_i$ over time $\Delta t$ we find the grid velocity
and the heat production by release of gravitational energy for each shell
(last term in Eq. \ref{EqT}).  

\item We calculate the pressures $p_i$ at shell boundaries from the discretised
pressure equation, starting with the given external pressure at the surface.
 
\item We solve for given temperatures $T_i$ and pressure $p_i$ at each
grid-point Eq. (\ref{DGLforD}) for the porosity over time interval $\Delta t$
to determine an updated value of $\phi_i$ at $t_k$. The corresponding 
non-linear equations are solved iteratively with an accuracy of better than
$10^{-12}$.  

\item We calculate from the updated porosity and pressure the heat conductivity.

\item We calculate for given $T_i$ the heat capacity.

\item The surface temperature $T_{\rm s}$ is determined from Eq. 
(\ref{BoundCondTsNeum}). This equation is solved for $T_{\rm s}$, using on the
l.h.s. the values for $T_i$ and $K$ from the last iteration step.

\item Updated values of temperatures $T_i$ at $t_k$ are calculated for all $i$
from the difference equations resulting from the heat conduction equation.

\item Check, if deviation of new values of $T_i$ and $\phi_i$ at $t_k$ from
current values is sufficiently small. 

\item If not, repeat calculation from step 3 on.

\item Otherwise determine new stepwidth $\Delta t$ (see below), advance $k$ by
one, and repeat from step 1 on for the next time step.

\end{enumerate}  
This kind of iteration converges usually within about ten to twenty steps. The
accuracy requirement was that the relative change of $T_i$, $\phi_i$ between
subsequent iteration steps is less than $10^{-8}$. This simple scheme works 
efficiently since the coefficients in the heat conduction equation do not
strongly depend on temperature. Then this equation can be solved by the simple
fixed-point iteration described. Test calculations done with complete
linearisation of the non-linear equation showed that this did not significantly
improved the efficiency of the solution method in our particular case. The 
advantage of our method is that it poses less stringent requirements on the
existence and continuity of derivatives than the Newton-Raphson method for
convergence of the iteration method.
 
The boundary condition given by Eq. (\ref{BoundCondTsNeum}) in principle should
be built into the difference equations for the heat conduction equation. 
Numerical experience showed that this occasionally resulted in stability
problems. Our present method is to solve Eq. (\ref{BoundCondTsNeum}) as a
separate equation using at the current iteration step a value for the conductive
heat flux at the surface calculated from the quantities of the last iteration
step. The resulting value of $T_{\rm s }$ is prescribed as Dirichlet boundary
condition, Eq. (\ref{HeatBoundSurfT}), for Eq. (\ref{EqT}). The temperatures $T_{\rm s}$ calculated this way at each iteration step converge to the solution
of Eq.~(\ref{EqT}) subject to Neumann boundary condition 
Eq.~(\ref{BoundCondTsNeum}). This method worked without problems.

The time steps $\Delta t$ are chosen such that the relative change of the 
variables over $\Delta t$ is about 3\%. This is sufficiently small that a 
further reduction of the stepwidth does not significantly improve the
accuracy of the numerical solution; a reduction of the admitted stepwidth 
by a factor of two results in our case in relative changes of the numerical 
values of the variables by a few times of $10^{-4}$. If the number of iteration
steps becomes too big (e.g. $>20$) with this choice, the stepwidth $\Delta t$ is
reduced by a factor of two until the number of iteration steps does no more
exceed the limit. Since we use an implicit solution method, there is no 
limitation for the stepwidth from stability requirements.

The initial model for the start-up of the solution method assumes a fixed
temperature ($=T_{\rm s}$ at initial time) within the body. An appropriate
set of masses $\Delta M_i$ is chosen that results (i) in a sufficiently fine
grid at the surface to resolve the rapid temperature variations at the surface
and that (ii) is sufficiently fine for allowing to calculate the derivative 
$\partial T/\partial M_r$ at the centre with sufficient accuracy. For the
initial model the porosities $\phi_i$ and radii $r_i$ for the set of mass-shells
are calculated from hydrostatic equilibrium and the equation of state for cold
pressing, as  described in Sect.~\ref{SectColdPress}.

\begin{figure}

\centerline{
\includegraphics[width=0.9\hsize]{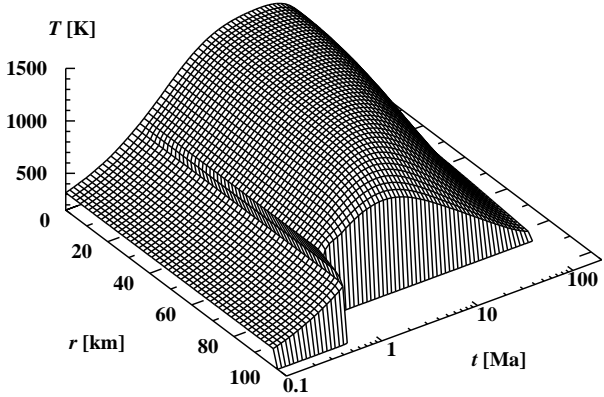}
}

\caption{Evolution of radial distribution of temperature for model PL (see
Table \ref{TabMiyaParm} for its definition). Note the radial shrinking of the
body by compaction of the initially porous material at about 0.6 decay 
timescales of $^{26}$Al.}

\label{FigTempEvol} 

\end{figure}

If a fixed temperature $T_{\rm b}$ is to be prescribed at the outer boundary,
this is technically achieved within the frame of our solution algorithm by
letting $F_{\rm ext}=\sigma T_{\rm b}^4$ in Eq.~(\ref{BoundCondTsNeum}).

The solution method also allows to consider growing bodies by increasing the
mass of the outermost shell according to some prescribed growth-law and
splitting this shell into two shells at each instant where its mass exceeds
some threshold value. This option in our code is not used, however, in the
model calculations presented in this paper. 
    
\subsection{Some sample calculations}

\subsubsection{The model of Miyamoto et al.}

As a first test we calculate with the code a model using the same model
parameters as in \citet{Miy81}. The basic parameters of the model are given in
Table  \ref{TabMiyaParm} in the column marked with MFT81. The model of 
\citet{Miy81} is one for a completely homogeneous body and does not consider
the effects of porosity and the possibility of sintering. The data assumed for
$K$ and $\varrho$ correspond to average properties of L chondrites that, in
fact, have low but non-vanishing porosities, scattering around about $\phi=10$\%
\citep{Yom83}. The true bulk density and heat conductivity of completely
consolidated chondritic material is higher ($\varrho_{\rm bulk}=3.6$\,g\,
cm$^ {-3}$, $K=4.27$\,W\,kg$^ {-1}$\,K$^ {-1}$, see \citet{Yom83}, their 
Table 5). The model is calculated by choosing as initial value $\phi=0$ which 
guaranties that during the course of the calculation the porosity remains zero.
Heating is only by decay of $^{26\!}$Al. The result for the temperature
evolution in the centre of the body and a number of selected radii is shown in
Fig.~\ref{FigTmevolMiya}a. This is almost identical with the result obtained by
\citet{Miy81} from the analytic solution of the heat conduction equation, i.e.,
our code reproduces this exact analytic result.

\subsubsection{Model of a porous body}

In Fig.~\ref{FigTmevolMiya}b we show the results for the temperature evolution
of a body having the same size and using a similar set of parameters, but now
considering that the heat conductivity of the porous material,
Eq.~(\ref{FitToKpor}), is different from the value of heat conductivity used by
\citet{Miy81}, and that the material from which the body forms is initially porous and
consolidates by sintering. The parameters of the model are given in Table 
\ref{TabMiyaParm} in the column marked with PL0.

\begin{figure}

\centerline{
\includegraphics[width=0.9\hsize]{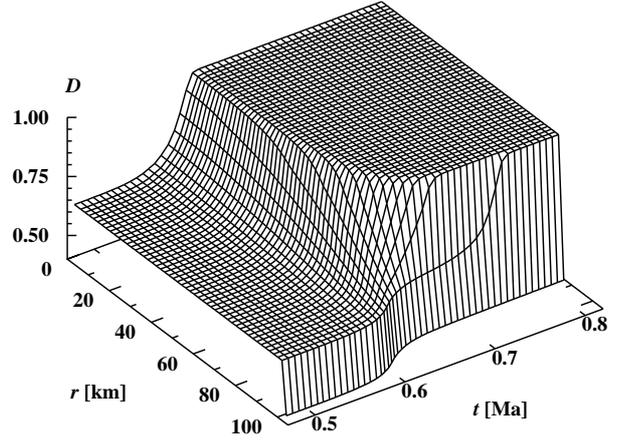}
}

\caption{Evolution of radial distribution of filling factor $D$ (relative
density) for model PL (see Table \ref{TabMiyaParm} for its definition). Shown
is the very initial phase of the evolution where the initially porous material
is compacted by sintering. The resulting shrinking of the planetesimal size
occurs at about 0.6 decay timescales of $^{26}$Al.}

\label{FigPorEvol} 

\end{figure}

It is assumed that the porosity of the surface layers at low pressures is 
$\phi_{\rm srf}=0.6$, corresponding to the degree of compaction found in 
\citet{Wei09} for powder material that was subject to numerous impacts. This is
what one expects for the early formation time of asteroids where they grow by
repeated slow impacts of much smaller bodies. In deeper layers of the body
where pressures are high due to self-gravity the material is compressed by
isostatic pressing to higher densities up to a limiting value of $\phi\approx
0.4$, see Sect.~\ref{SectColdPress}. The corresponding initial distribution of
porosities in the interior is calculated as described in 
Sect~\ref{SectColdPress}. For typical results see Fig.~\ref{FigPhiPIso}. This
kind of compaction is a purely mechanical effect due to mutual rolling and
gliding of the powder particles driven by an applied pressure which requires no
elevated temperatures and acts therefore already in cold bodies (cold pressing).
Starting from this initial configuration the evolution of the model was
calculated. The porosity dependence of the heat conduction is taken into account
by means of approximation Eq.~(\ref{FitToKpor}). The surface temperature was
taken to be constant over time and equal to $T_{\rm s}=150$\,K. 

\begin{figure}

\centerline{
\includegraphics[width=.48\hsize]{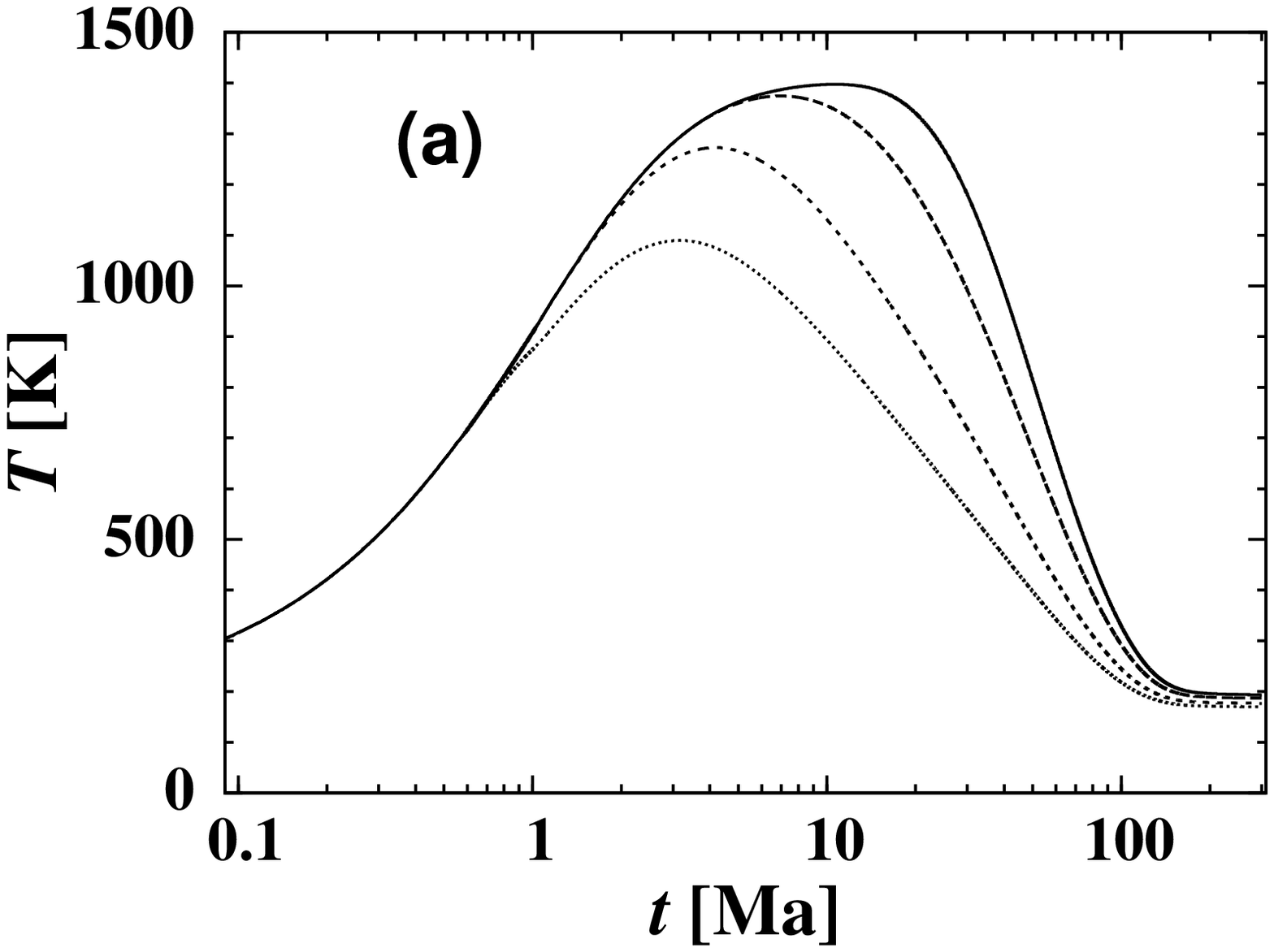}
\hfill
\includegraphics[width=.48\hsize]{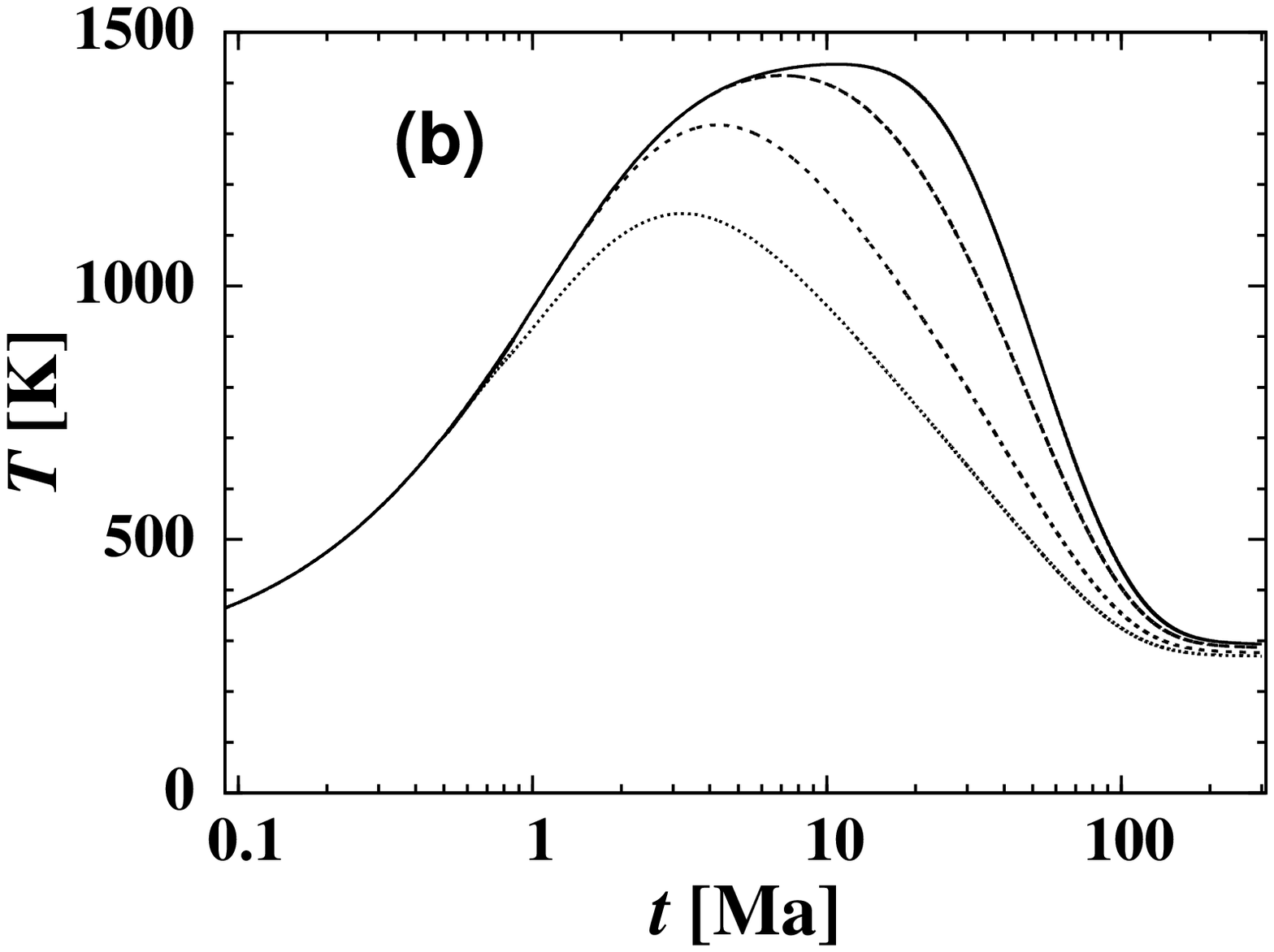}
}

\centerline{
\includegraphics[width=.48\hsize]{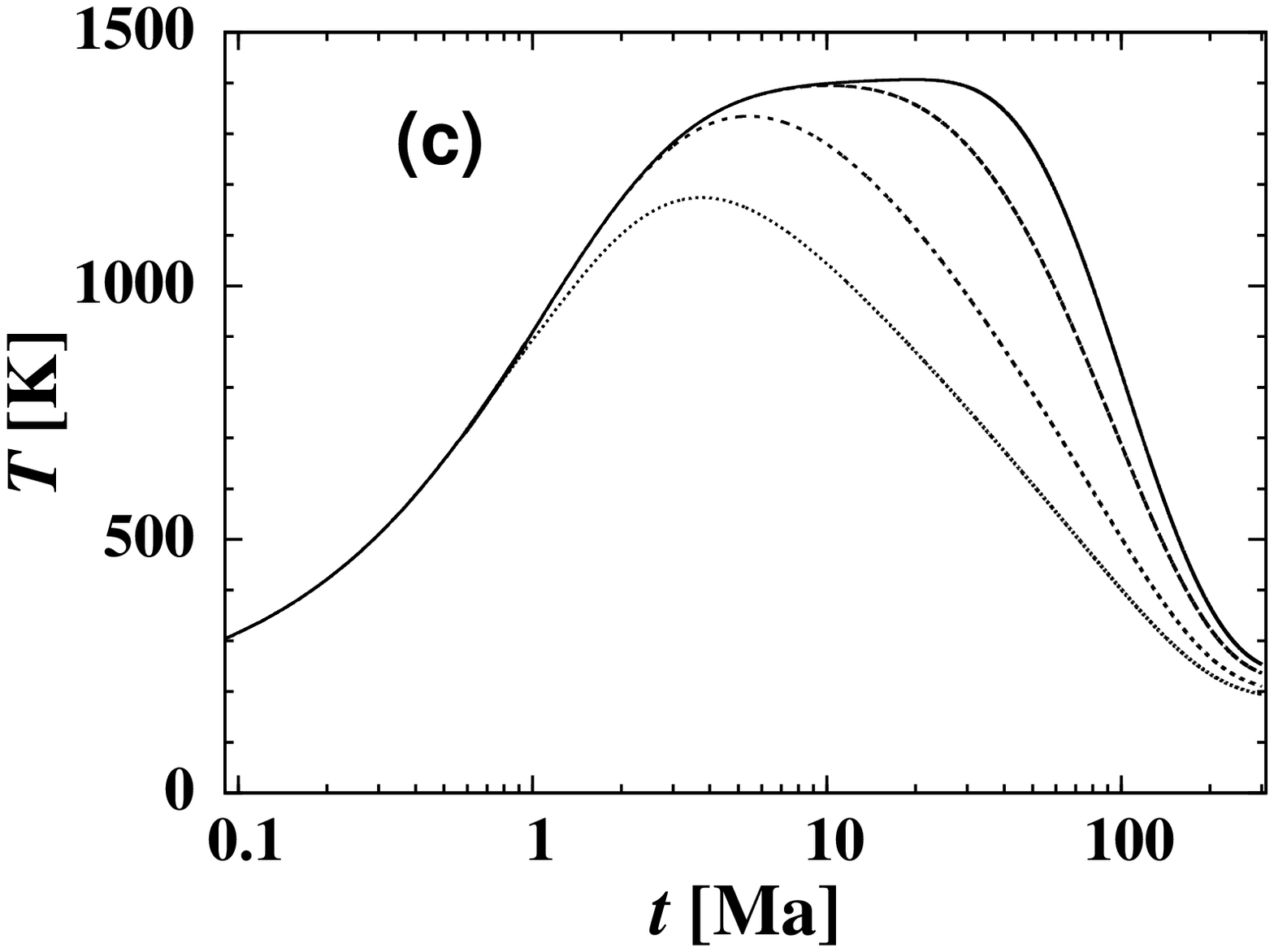}
\hfill
\includegraphics[width=.48\hsize]{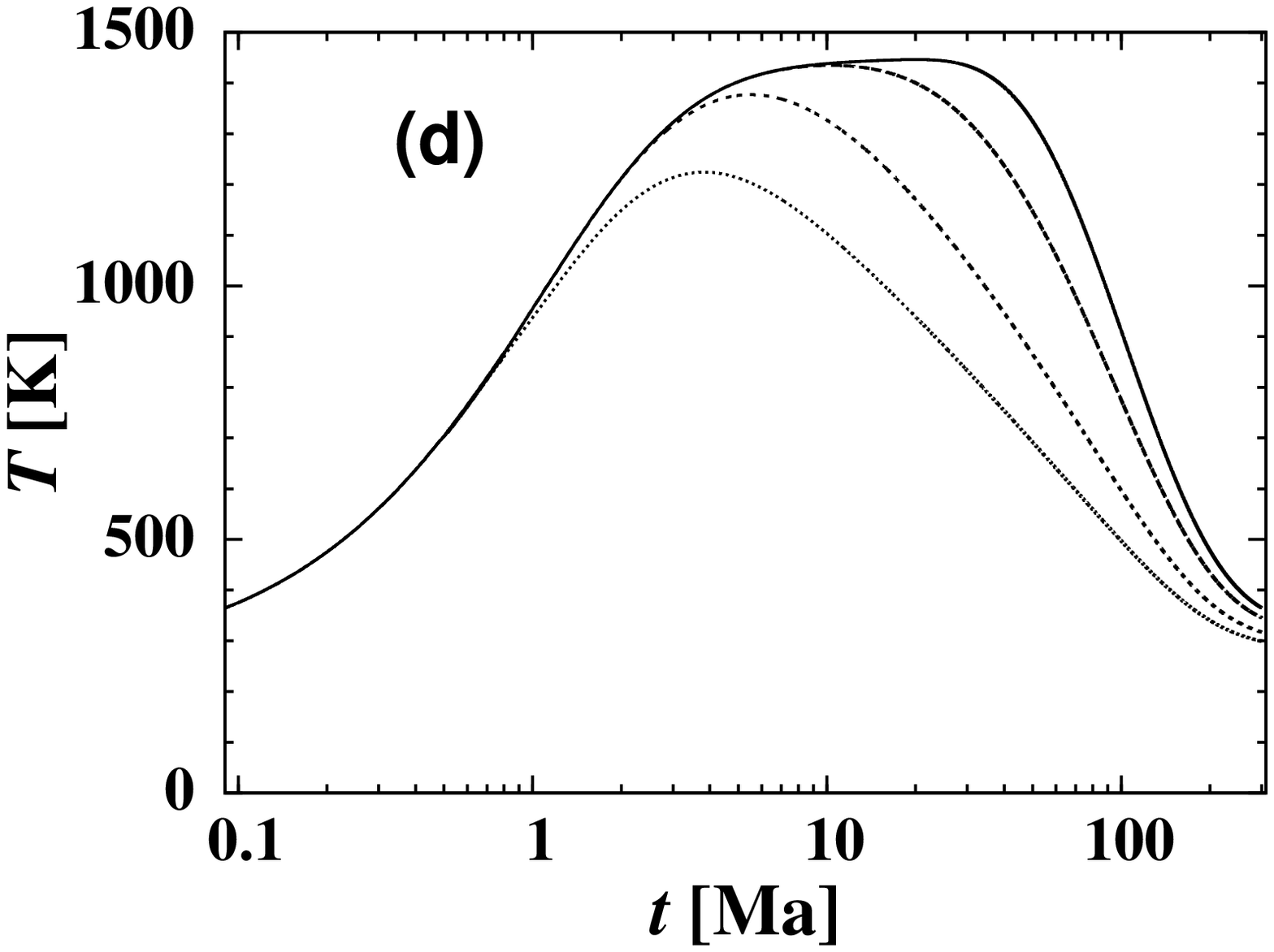}
}

\caption{Temperature evolution of test models for a porous bodies with modified
values of parameters. Left panel with boundary temperature $T_{\rm b}=150$\,K,
right panel with $T_{\rm b}=250$\,K. Upper panel with heat conduction 
coefficient $K_{\rm b}=4$\,W\,m$^{-1}$\,K$^{-1}$, lower panel with 
$K_{\rm b}=2$\,W\,m$^{-1}$\,K$^{-1}$.
}

\label{FigVarTK} 

\end{figure}

Figure \ref{FigTmevolMiya}b shows that the peak temperature achieved in the
centre of the body is lower in this model than in the model of \citet{Miy81}.
This results from the higher heat conductivity in our model after complete
sintering ($K_{\rm b}=4$ vs. $K_{\rm b}=1$). Contrary to this,
the temperature in the outer layers of
the model increases more rapidly than in the model of \citet{Miy81} because the
high porosity outer layer acts as an insulating shield that prevents an 
efficient loss of heat to outer space. At a temperature of about 700--750\,K
(depending on pressure) sintering by hot pressing becomes active and the
porosity rapidly approaches $\phi=0$ at higher temperatures. The temperature
structure then becomes nearly isothermal in the interior of the body and the 
temperature drops to the outer boundary value within a layer of only a
few km thickness. This is shown in detail in Fig.~\ref{FigTempEvol}. 

Figure~\ref{FigPorEvol} shows the evolution of the porosity during the earliest
evolutionary phase. An outer shielding powder layer persists during the whole
evolution of the body because cooling of the outer layers prevents the material
to warm-up to the threshold temperature at about 750\,K for compaction by hot
pressing at low pressures. This behaviour is completely analogous to what is
found in \citet{Yom84} and can be compared with results by
\citet{Sah07} and \citet{Gup10}. They considered gradual sintering in the
temperature range of 670--700 K by a smooth interpolation recipe, reducing 
the porosity from 55\% to 0\% by compaction of the body, and took into 
account respective changes in thermal diffusivity.

\begin{figure*}

\centerline{
\includegraphics[width=.32\hsize,angle=-90]{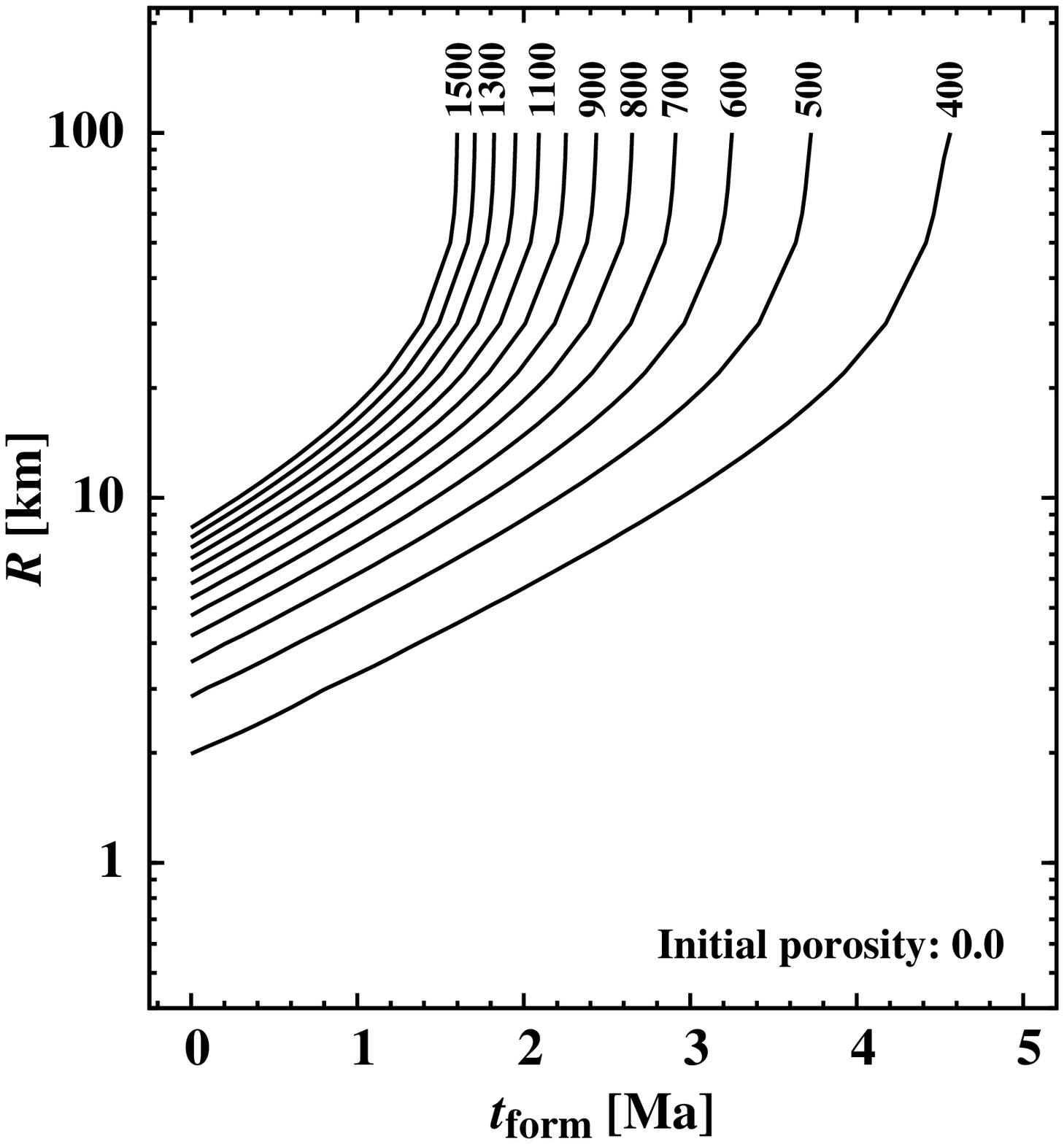}
\hfill
\includegraphics[width=.32\hsize,angle=-90]{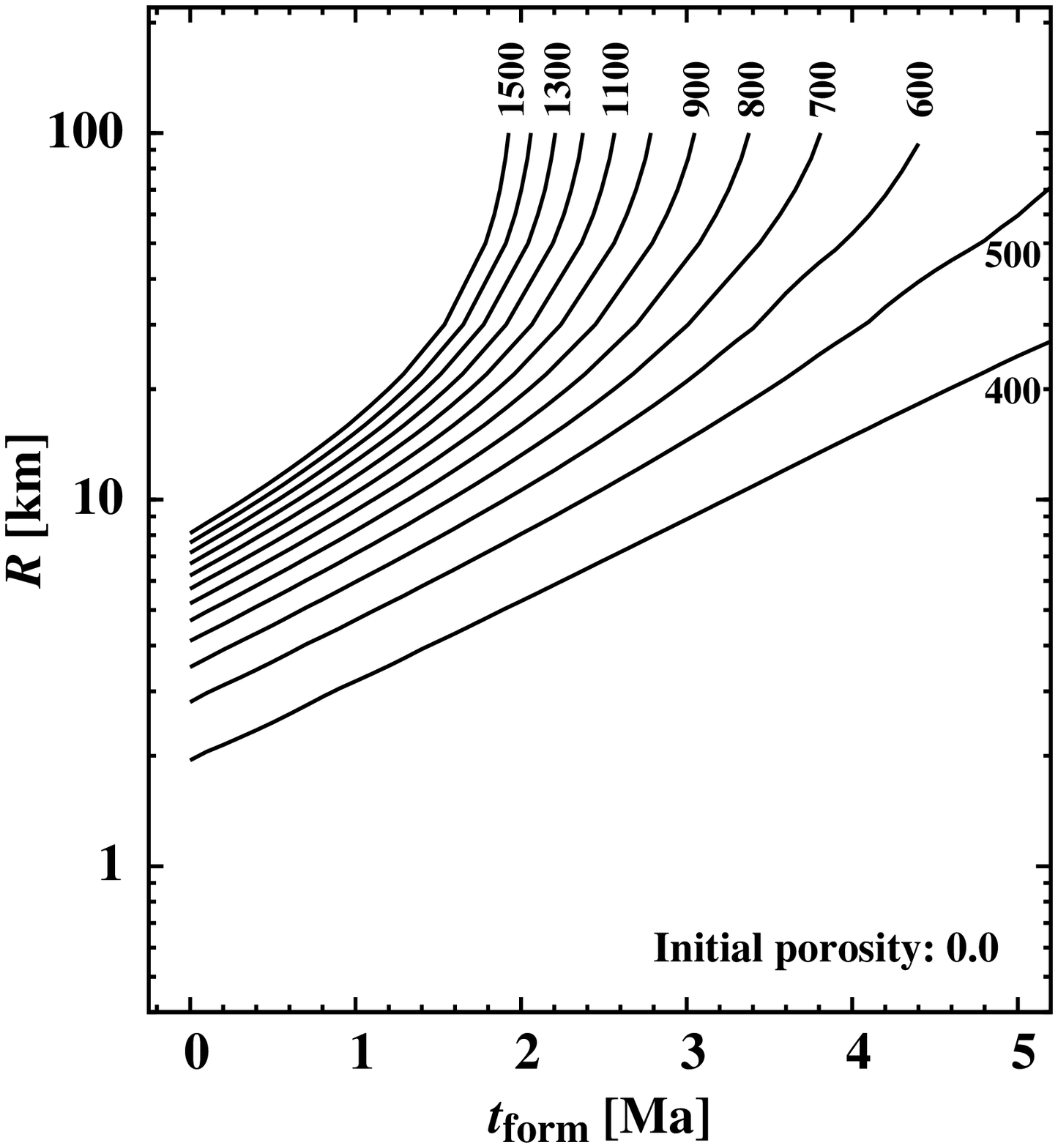}
\hfill
\includegraphics[width=.32\hsize,angle=-90]{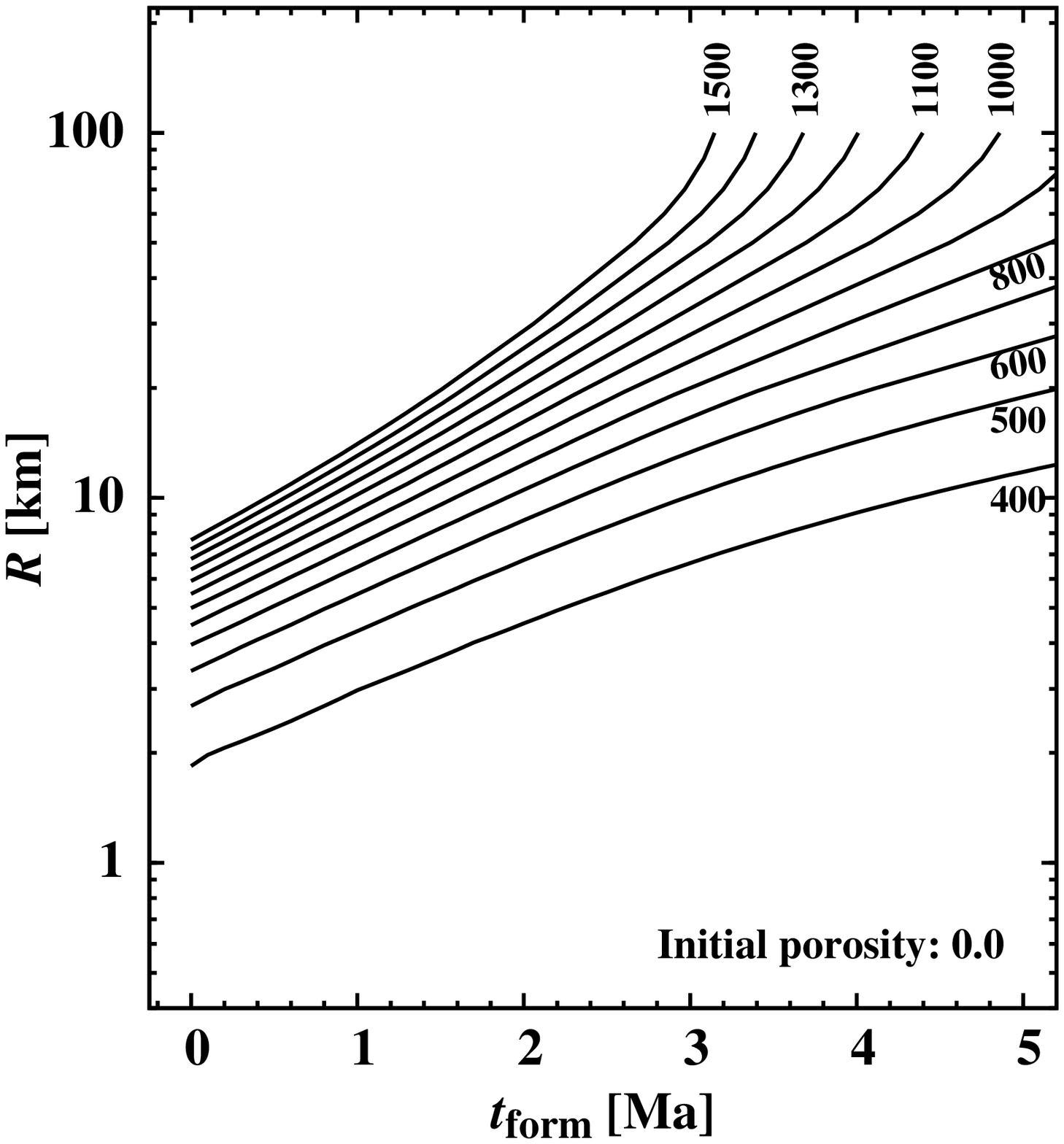}
}

\centerline{
\includegraphics[width=.32\hsize,angle=-90]{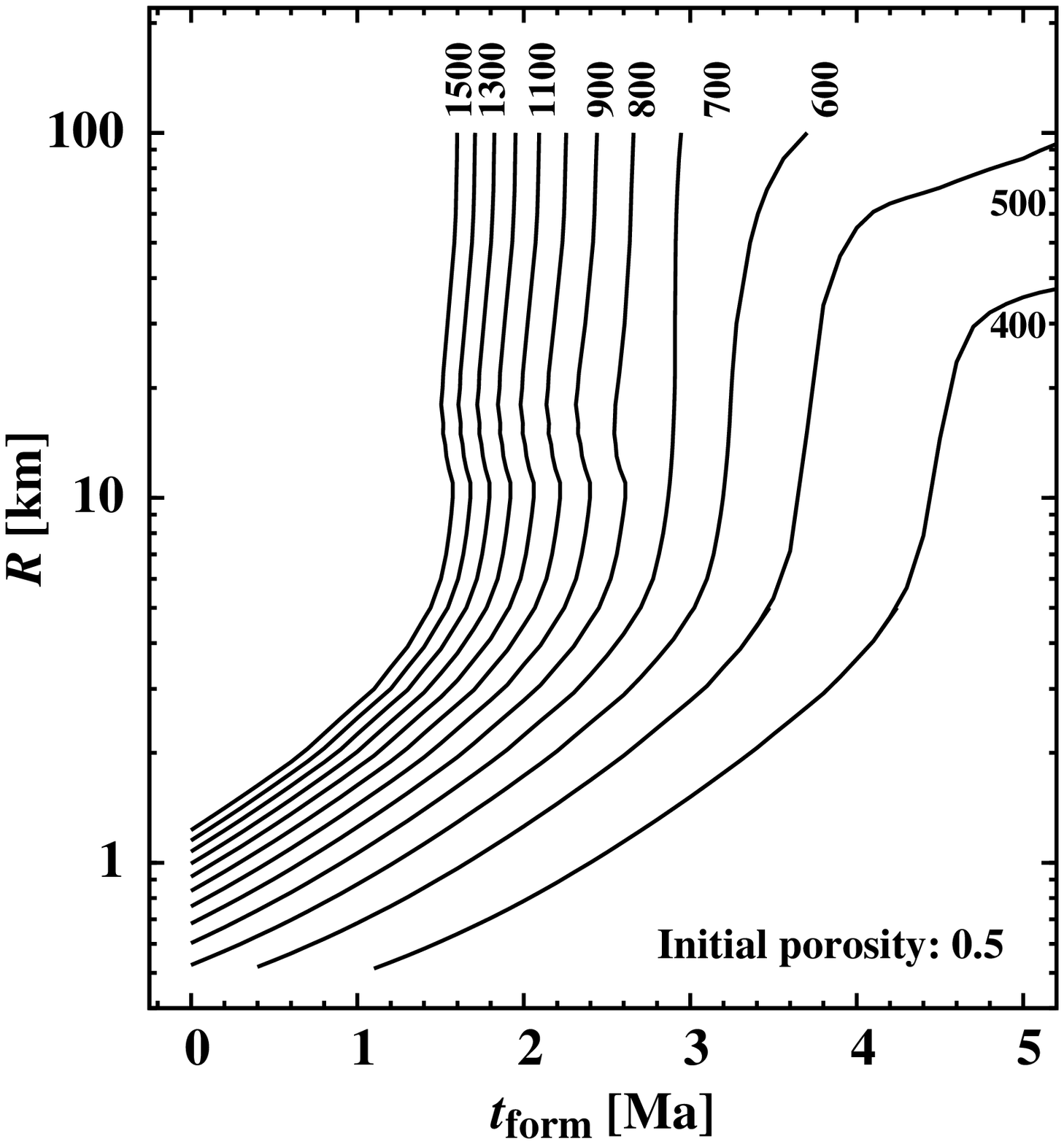}
\hfill
\includegraphics[width=.32\hsize,angle=-90]{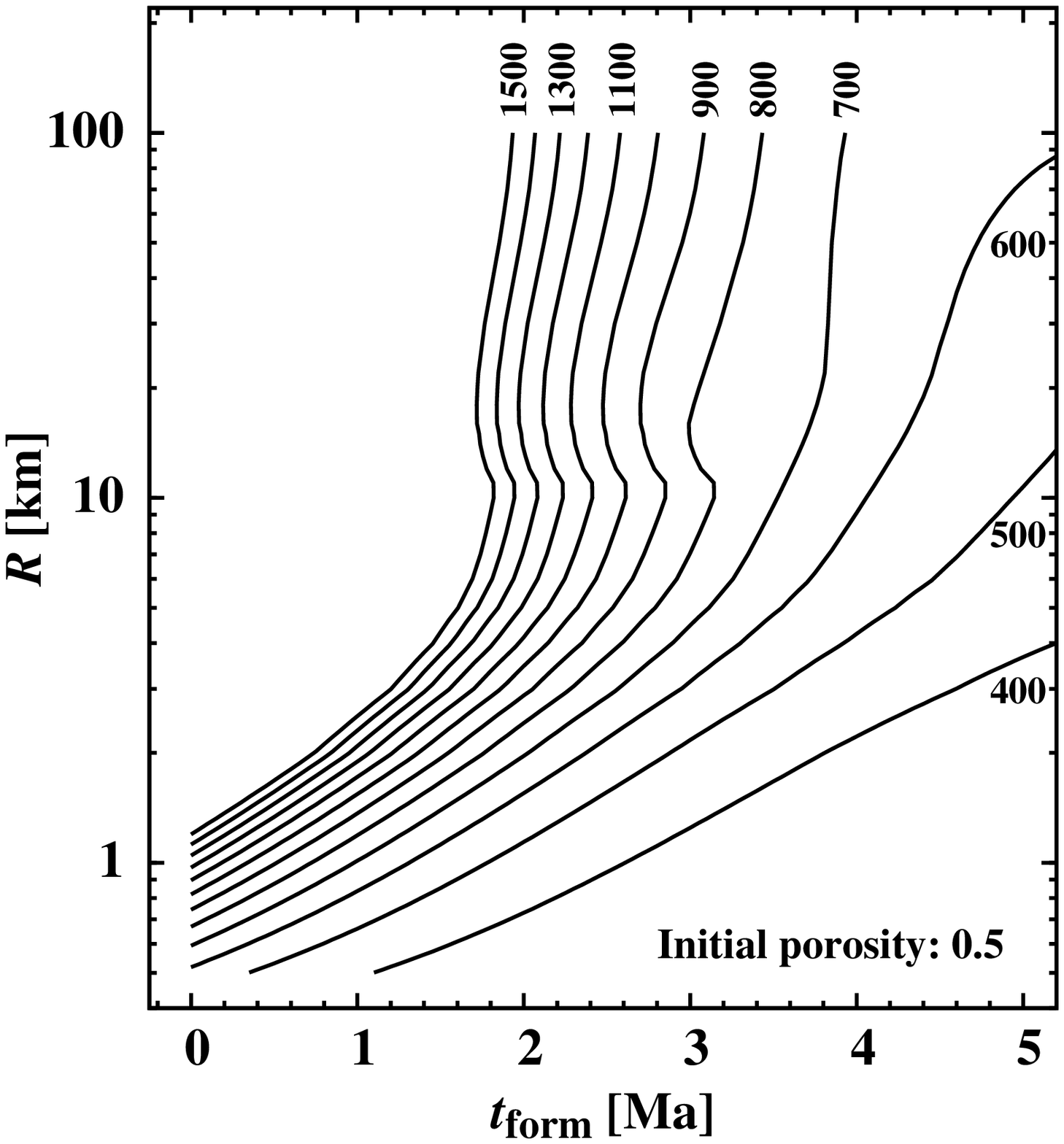}
\hfill
\includegraphics[width=.32\hsize,angle=-90]{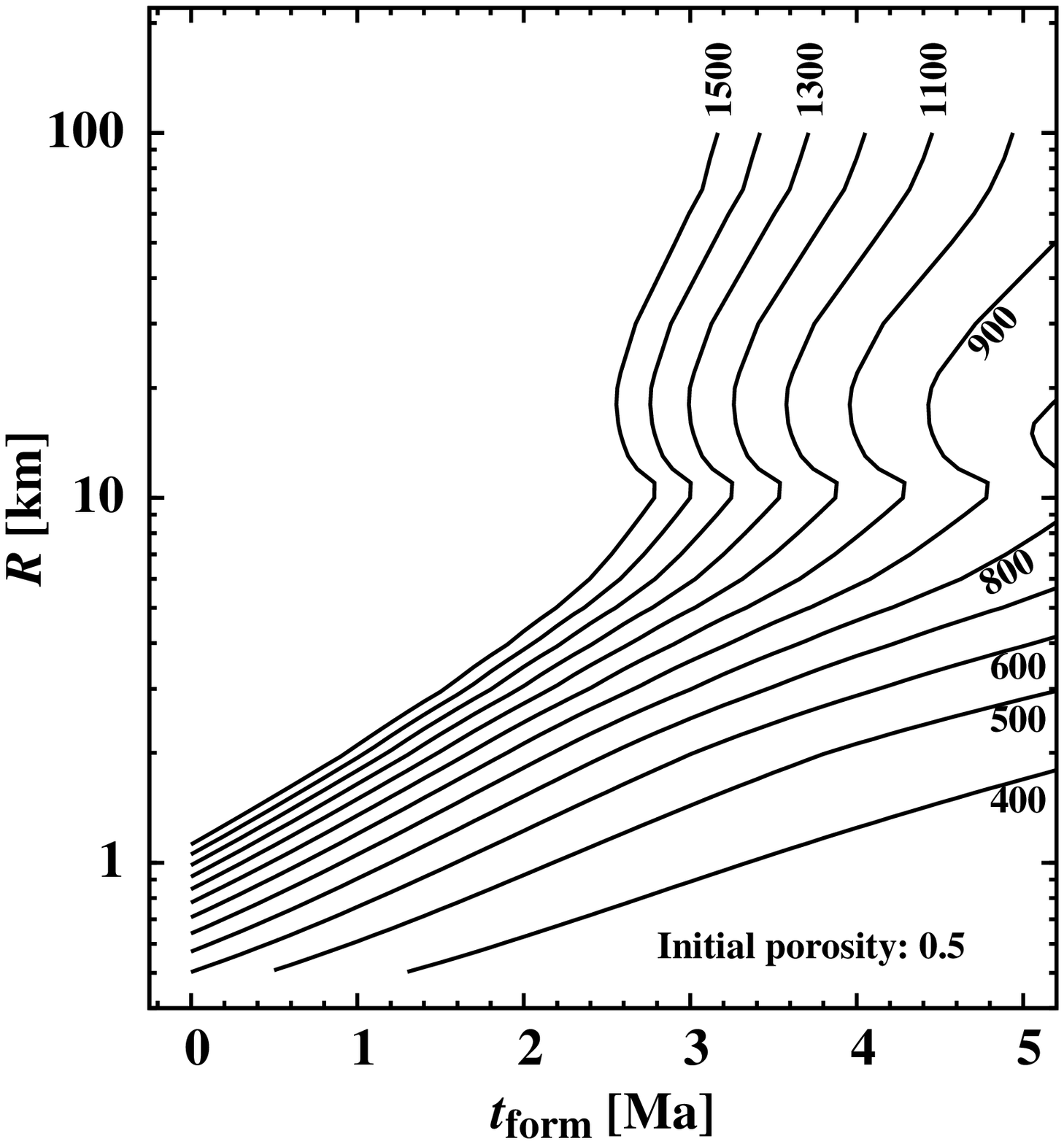}
}

\caption{Variation of maximum temperature in the centre of a planetesimal with
radius $R$ and instant of formation $t_\mathrm{form}$. \textit{Upper panel:}
Completely consolidated bodies with porosity $\phi=0$. \textit{Lower panel:}
Models of initially porous bodies with porosity $\phi_{\rm srf}=0.5$ which
consolidate in the interior during the course of their thermal evolution. The
lines correspond to the indicated maximum central temperature. At temperatures
above $\approx 1\,400$\,K partial melting of the mineral mixture is expected.
Temperatures exceeding 1\,500,K are left out for this reason. Shown are models
for three different initial 
$^{60}$Fe/$^{56}$Fe abundance ratios: \textit{Left panel:} Without $^{60}$Fe.
\textit{Middle panel:} The optimum fit value of $4.1\times10^{-7}$ for the H
chondrite parent body, see Sect.~\ref{SectFitHChond}. \textit{Right panel:}
Highest value of $1.6\times10^{-6}$ given in Table ~\ref{TabMiyaParm}.
}

\label{FigIsoth} 

\end{figure*}

Because the porosity approaches zero everywhere in the body where the
temperature exceeds the threshold for sintering by hot pressing, the radius of
the body shrinks significantly, typically by 20\% of its initial value. This
occurs after about 0.6 decay timescales of the main heat source, $^{26\!}$Al, in
this model, cf. Figs.~\ref{FigTempEvol} and \ref{FigPorEvol}. The size of the
model, marked as PL0 in Table~\ref{TabMiyaParm}, of 85\,km corresponds to the
final radius that the body would have after compaction to zero porosity (the
initial radius before sintering is $\approx\,105$\,km). The final radius
of the body almost exactly equals the hypothetical final radius at zero
porosity, since the powder layer remaining at the surface is rather thin. Also
the temperatures shown in Fig.~\ref{FigTmevolMiya}b for a number of depth points
below the surface correspond to that Lagrangean  $M_r$-coordinate, which after
compression to zero porosity would have the given value of the radius
coordinate. Before the body shrinks these points have somewhat bigger depths
below the surface.      
  
In Fig.~\ref{FigTmevolMiya}c we show the temperature evolution in a model with
the same set of parameters as the previous model (see Table~\ref{TabMiyaParm},
model PL), that considers heating by decay of $^{60}$Fe and long-lived
radioactive nuclei as additional heat sources, using an $^{60}$Fe/$^{56}$Fe
ratio at the upper limit of observed values (see Sect.~\ref{SectHeatRate}). 
The peak central temperature is about 30\% higher than without this heat source.

\subsubsection{Mass-shells and time steps}

Since in the present models the mass of the body is constant, a fixed grid of 
mass-shells is used in the calculations. This grid was constructed as follows:
Starting from an outer layer with some (small) thickness, the thickness of the
layers from the outside to the interior increases by a constant factor from one
shell to next such that for some given number $K$ of mass-shells the radius
of the innermost sphere is 100\,m; this fixes the width of the outermost layer.
The models of this paper are calculated with $K=300$, in which case the 
outermost layer has a thickness of $\approx3$\,m. This choice of grid guarantees
a sufficient high resolution of the thin outer region of rapid drop of 
temperature. An increase of $K$ to $600$ does not result in significant changes
of the model characteristics; the relative change of central temperature, e.g.,
is $\approx 10^{-4}$.
 
In the model calculations used for fitting models to empirical cooling
histories of meteorites described in Sect.~\ref{SectFitHChond}, the total number
$K$ of shells is increased to $K=1200$. This is necessary if one requires
that even in the region of steepest temperature decrease toward the surface the
relative changes of temperature at some fixed mass-coordinate are at most of the
order of a few times $10^{-4}$ if the number of grid points is doubled.

The timesteps $\Delta t$ are chosen according to the strategy described in 
Sect.~\ref{SectMethModelCalc}. The timesteps during the initial heat-up phase
were typically a few thousand years. Once sintering commences, the step size 
decreases to about 100 years and varies around this value until complete
sintering of the body (except for the outermost layers), Then $\Delta t$
increases continuously and during the final phase is limited to a maximum value
of 1\,Ma in order to obtain not too widely spaced steps for plotting purposes.
The total number of timesteps required for a complete model run for an evolution
period of 100\,Ma is between 3\,500 and 4\,000.

\subsubsection{Dependence on parameters}

In order to get an impression how the model depends on some important 
parameters, we show in Fig.~\ref{FigVarTK} results for the same kind of model
as model PL, but calculated with two different values of the surface temperature
$T_{\rm s}$ and bulk heat conductivity coefficient $K_{\rm b}$. The models in
the left panel are calculated with a fixed surface temperature of 150\,K, the
models in the right panel with a fixed surface temperature of 250\,K. These two
values encompass the possible values of the surface temperature for bodies in
the region of the asteroid belt for both cases, if either the body is embedded
in the accretion disc (cf.  Fig.~\ref{FigPhiPIso}) or the accretion disc has
gone and the body is under the influence of the radiation of the young sun.
There are only small differences between the run of the temperature at different
depths, i.e., the temperature evolution below the immediate surface layer does
not critically depend on the surface temperature, at least not for bodies with
radii of the order of 100\,km which are our main topic. Therefore we do not
attempt in our further calculations to calculate $T_{\rm s}$ as precise as
possible from  Eq.~(\ref{BoundCondTsNeum}) and simply assume a reasonable but
fixed value for all times.
  
The upper panel in Fig.~\ref{FigVarTK} is calculated with a value for the heat
conduction coefficient of $K_{\rm b}=4$\,W\,m$^{-1}$\,K$^{-1}$, the lower panel 
with $K_{\rm b}=2$\,W\,m$^{-1}$\,K$^{-1}$. The first value corresponds to what
has been found as the average value for H and L chondrites if measured values 
are extrapolated to zero porosity, see Sect.~\ref{SectApptHeatK}. As one
can see from Fig.~\ref{FigTvarKb} there is significant scatter in the heat
conduction coefficient (of presently unknown origin) and it is not clear
whether the investigated sample of H and L chondrites are representative for
the whole material of the parent body of the H or L chondrites. The value of  
$K_{\rm b}=4$\,W\,m$^{-1}$\,K$^{-1}$ corresponds to typical values of $K$ for
pure silicate minerals \citep[cf.][]{Yom83} and therefore probably represents
the upper limit for the possible values of $K_{\rm b}$. Lower values,
therefore, may also be of interest for real planetesimals. Figure~\ref{FigVarTK}
shows that the results of the model calculation depend significantly on the
value of $K_{\rm b}$. Because it is presently not possible to determine 
$K_{\rm b}$ from first principles for chondritic material, we consider 
$K_{\rm b}$ in our later model calculations to be a free parameter (but, of
course, restricted to the range of values found for chondrites).

\subsection{Maximum central temperatures}

The maximum central temperature that is reached during the course of the
evolution of a planetesimal is an indicator of what kind of changes the material
may undergo. If the central temperatures exceed the solidus temperature of 
chondritic material of about 1\,400\,K \citep{Age95} partial melting occurs and
the body starts to differentiate. If the temperature stays below the threshold
temperature of about 700\,K (at $\approx0.1$\,bar) for sintering, the whole body
retains its porous structure. The maximum central temperature T$_{c,\rm max}$
depends mainly
\begin{enumerate}
\item
on the radius, $R$, of the body, because this determines how efficiently heat
can be removed from the central region by heat conduction, and 
\item
on the formation time, $t_{\rm form}$, because this determines how much of the
initial inventory of radioactive material already decayed before the body grew
to its final size.
\end{enumerate}
Figure~\ref{FigIsoth} shows the dependence of $T_{c,\rm max}$ on 
$R$ and $t_{\rm form}$ for models of initially porous bodies and, for
comparison, of bodies with pore-free material. Obviously the thermal evolution
history of initially porous bodies is very different from history of equal
sized compact bodies. Models are shown for three different assumptions on the
abundance of $^{60}$Fe as additional heat source besides $^ {26\!}$Al.

\begin{table}

\caption{Key data for cooling history of selected H chondrites}

\begin{tabular}{l@{\hspace{1cm}}lll}
\hline\hline
\noalign{\smallskip}
Quantity & Kernouv\'e & Richardton \\
\noalign{\smallskip}
\hline
\noalign{\smallskip}
type    & H6 & H5 \\
\noalign{\smallskip}
         & \multicolumn{2}{c}{Hf-W$^3$ (metal-silicate)} \\
\noalign{\smallskip}
radiometric age & 4559$\pm$1 & 4563$\pm$1 & Ma\\
temperature     & 1150$\pm$75 & 1100$\pm$75 & K \\
\noalign{\smallskip}
         & \multicolumn{2}{c}{U-Pb-Pb$^3$ (phosphates)} \\
\noalign{\smallskip}
radiometric age & 4522$\pm$1 & 4551$\pm$1 & Ma\\
temperature     & 720$\pm$50 & 720$\pm$50 & K \\
\noalign{\smallskip}
         & \multicolumn{2}{c}{Ar-Ar$^{2,3}$ (feldspar)} \\
\noalign{\smallskip}
radiometric age & 4499$\pm$6 & 4525$\pm$11 & Ma\\
temperature     & 550$\pm$20 & 550$\pm$20 & K \\
\noalign{\smallskip}
         & \multicolumn{3}{l}{Pu-fission tracks (pyroxene-merrilite)} \\
\noalign{\smallskip}
calculated age$^4$ & 4499$\pm$6 & 4525$\pm$11 & Ma \\
cooling rate & 2.6$\pm$0.5 & 2.9$\pm$0.5 & K/Ma \\
\noalign{\smallskip}
calculated age$^4$ & 4438$\pm$10 & 4469$\pm$14 & Ma \\
time interval$^{1,3}$ & 61$\pm$8 & 56$\pm$9 & Ma \\
\noalign{\smallskip}
         & \multicolumn{2}{c}{metallographic } \\
\noalign{\smallskip}
 cooling rate$^{1,3}$    & 10 & 20 & K/Ma \\
\noalign{\smallskip}
\hline
\end{tabular}

\medskip{\scriptsize Notes:
\\
$^1$~\begin{minipage}[t]{.975\hsize}
Time-interval for Pu-fission track cooling rate from 550--390\,K,
metallographic cooling rate 800--600\,K
\end{minipage}

$^2$~\begin{minipage}[t]{.975\hsize}
Recalculated for miscalibration of K decay constant \citep[explanation
see text and][]{Tri03}
\end{minipage}

$^3$~\begin{minipage}[t]{.975\hsize}
Data from Hf-W: \citet{Kle08}, U-Pb-Pb: \citep{Goe94}, Ar-Ar: \citet{Tri03},
metallographic cooling rates: \citet{Tay87}
\end{minipage}

$^4$~\begin{minipage}[t]{.975\hsize}
Calculated age at 390\,K from time interval between Pu-fission track (merrilite, 390\,K)
and Pu-fission tracks at 550\,K (pyroxene) compared to Ar-Ar feldspar age at
550\,K
\end{minipage}

}

\label{TabDatChondCool}
\end{table}

For porous bodies smaller than $\approx5$\,km radius the initial porosity is
very high because they are even not compacted by cold pressing (cf. 
Fig.~\ref{FigPhiPIso}). Because of their low initial heat conductivity even
rather small bodies ($R\gtrsim0.5$\,km) heat up at least to threshold 
temperature for sintering and become compacted in their interior, because the
strongly insulating powder layer on their surface prevents their cooling.
Completely compact bodies would reach central temperatures higher than 700\,K
only for radii exceeding $\approx5$\,km because of much more efficient heat
conduction.

\begin{figure*}
\sidecaption
\includegraphics[width=12cm]{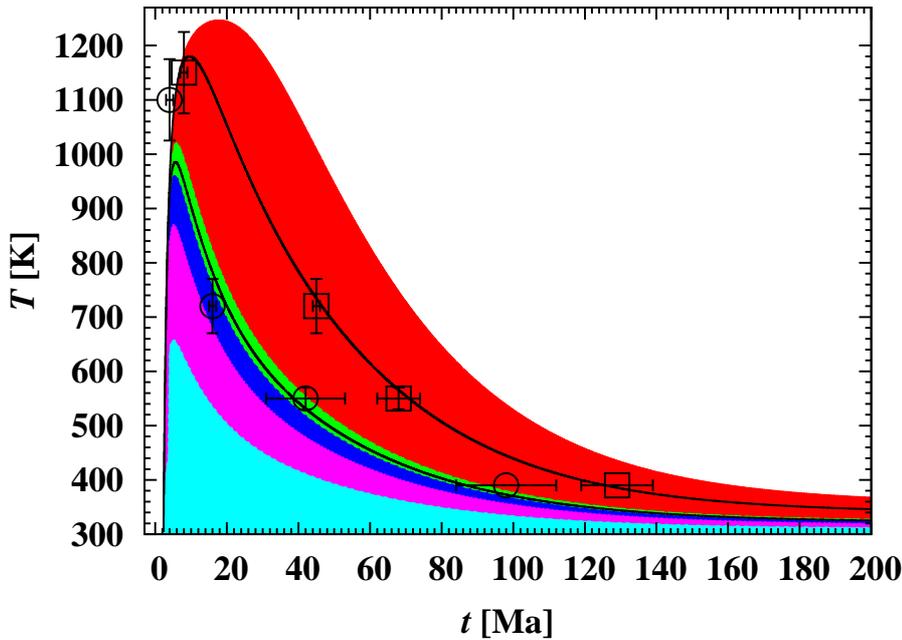}
\caption{Optimum-fit model for the cooling history of the parent body of H
chondrites. Abscissa is time elapsed since formation CAIs. Evolution of 
temperature at a number of depths below the surface is shown. The upper
contours of shaded areas correspond (from bottom to top) to depth's of
0.32\,km, 2.3\,km, 7.8\,km, 11\,km, and the uppermost contour to the centre.
The rectangular boxes and circles  correspond to the empirical data for H6 and
H5 chondrites, respectively, given in Table~\ref{TabDatChondCool}. Crosses are
error bars. The dashed lines correspond to the temperature evolution at depth's
of 8.9\,km (lower line) and 36\,km (upper line) below the surface. They
represent our best fit to the empirical data.
}

\label{FigOptFitH}
\end{figure*}

For initially porous bodies bigger than $\approx50$\,km radius already the
initial porosity is low throughout almost all of the body because such bodies
are already  strongly compacted by cold pressing (cf. Fig.~\ref{FigPhiPIso}) and
the remaining porosity rapidly disappears by sintering. Their thermal evolution
is essentially identical to that of completely compact bodies, except, of
course, in the layers close to the surface that retain part of their initial
porosity.
  
Porous bodies with radii between $\approx5$\,km and $\approx20$\,km are already
significantly compacted in their central part by cold pressing (cf. 
Fig.~\ref{FigPhiPIso}) but have initially an extended low-porosity outer mantle.
Porous bodies with radii between $\approx20$\,km and $\approx50$\,km also are
already compacted throughout the body by cold pressing  (cf. 
Fig.~\ref{FigPhiPIso}), except for the outer $\approx10$\% of their radius. They
show the most complex dependence of  $T_{c,\rm max}$ on $R$ and formation time.

Temperatures above $T_{c,\rm max}1\,500$\,K are not considered because we
consider in this paper parent bodies of undifferentiated meteorites. At a
temperature of $T_{c,\rm max}\gtrsim1\,400$\,K the silicates start melting 
\citep{Age95} and differentiation is unavoidable.

\begin{table}

\caption{Properties of optimum fit model}

\begin{tabular}{l@{\hspace{1cm}}lll}
\hline\hline
\noalign{\smallskip}
Quantity & & value & unit \\
\noalign{\smallskip}
\hline
\noalign{\smallskip}
Radius              & $R$              & 100 & km \\
formation time      & $t_{\rm form}$   & 2.3 & Ma \\
heat conductivity   & $K_{\rm b}$      & 4.0 & W\,m$^ {-1}$\,K$^{-1}$ \\
surface temperature & $T_{\rm s}$      & 300 & K \\
$^{60}$Fe/$^{56}$Fe abundance ratio &  & $4.1\times10^{-7}$ & \\
initial porosity    & $\phi_{\rm srf}$ & 50\% & \\
\noalign{\smallskip}
\hline
\noalign{\smallskip}
Chondrite type      & \multicolumn{2}{l}{depth-range} & \\
                    &  from & \hspace*{.3cm}to        & \\
\noalign{\smallskip}
\hline
\noalign{\smallskip}
powder layer        & 0.00  & \hspace*{.3cm}0.29     & km \\
H3                  & 0.29  & \hspace*{.3cm}2.3      & km \\
H4                  & 2.3   & \hspace*{.3cm}6.7      & km \\
H5                  & 6.7   & \hspace*{.3cm}10.8     & km \\
H6                  & 10.8  & \hspace*{.3cm}100      & km \\
\hline
\noalign{\smallskip}
                 &   & burial depth &   \\
\noalign{\smallskip}
\hline
\noalign{\smallskip}
Kernouv\'e       &   & 36           &  km \\
Richardton       &   & 8.3          &  km \\
\noalign{\smallskip}
\hline
\end{tabular}

\label{TabDatChondFit}
\end{table}

\section{%
Thermal history of the H chondrite parent body}
\label{SectFitHChond}

\subsection{Empirical data for cooling history}

Most meteorites reach the Earth as cm- or m-sized rocks, as they are the result
of repeated impact fragmentation of the initially much larger parent bodies. As
such events can be highly energetic, they change both chemistry, structure, and
also disturb the isotopic memory (i.e., the age information). Hence, 
information of cooling histories extending back to the origin of the solar
system must be restricted to meteorites that
\begin{enumerate}
\item show extraordinarily low mineralogical and structural
characteristics of shock metamorphism induced by asteroid collisions,
\item were dated with high precision, and
\item were dated simultaneously by a set of various high and 
low temperature chronometers tracing the cooling history from high temperatures
($>1\,000$\,K, e.g., Hf-W) down to intermediate (e.g. U-Pb-Pb or K-Ar ) and
very low temperatures (if possible $<400$\,K, e.g. $^{244}$Pu fission tracks).
\end{enumerate}
Such high quality data on un-shocked chondrites are quite limited, in spite of
the high  number of meteorites available in terrestrial collections.
While in the case of L chondrites, a major impact 470 Ma ago \citep{Kor07}
seems to have deeply erased the primordial low temperature cooling history,
H chondrites seem more promising. For a number of seven - noticeably 
un-shocked - H chondrites, complete high precision Hf-W, U-Pb-Pb, Ar-Ar and
$^{244}$Pu fission track data along with metallographic  cooling rate data
exist. Table~\ref{TabDatChondCool} shows such data for the H6 chondrite
Kernouv\`e and the H5 chondrite Richardton, which we can use for a preliminary
sample calculation.

\subsection{Fit of selected H chondrites}

A "fit" to the data in Table~\ref{TabDatChondCool} is shown in 
Fig.~\ref{FigOptFitH}. The chronological data for these two meteorites best fit
cooling curves in an asteroid at 36 and 8.9 km depth. The properties of the
parent body in Table~\ref{TabDatChondFit} were obtained according to the
following fit-procedure: The initial abundance of $^{26\!}$Al is roughly
constrained by the $^{26\!}$Al/$^{27}$Al ratio of ordinary chondrite chondrules,
which is an upper limit
shortly before parent body formation. Chondrule measurements indicate an 
$^{26\!}$Al abundance corresponding roughly to 2-3 Ma formation time after CAIs. 
Similarly, the $^{60}$Fe abundance is constrained by primitive type~3 ordinary
chondrites \citep[sulphides, see][]{Tac03}. Furthermore, abundances of
$^{26\!}$Al and $^{60}$Fe (or, in other words, the formation time of the parent
body) must be such that sufficiently high maximum metamorphic temperatures in
the asteroid are achieved to allow strong type 6 metamorphism, but also not too
high to prevent partial melting, metal silicate separation and differentiation.
Heat conductivity and radius mainly influence the total duration of the cooling
time of the parent body. We arbitrarily chose 100\,km, although a smaller 
asteroid would also allow extended cooling as observed for Kernouv\`e, which in
this model  needs only a burial depth of about 36 km. The boundaries separating
H6 from H5, H4 and H3  material are relatively shallow in the model, due to the 
insulating porous thin outer layer.

\subsection{Discussion}

The most prominent feature in our new model is the possibility of relatively
small parent planetesimals with significant heat retention. In the H chondrite
parent body model, this shows up in relatively thin layers of less heated or
metamorphosed material. 
Moreover, the relatively fast cooling required to
achieve temperatures below Hf-W and U-Pb-Pb closure for the H5 chondrite
Richardton (within 3 and 13 Ma, respectively) sets an upper limit to the
contribution of decay heat of $^{60}$Fe (roughly about 20-30\%). For example,
if $^{60}$Fe contributed all decay heat of ordinary chondrite parent body
metamorphism, sufficient fast cooling of H5 metamorphosed material would not
have been possible, as the half -live of $^{60}$Fe is 2.6\,Ma (a new revised
value) versus 0.72\,Ma for $^{26\!}$Al, implying significantly longer heat
production and implicit slower cooling in the first few Ma. This result is
in line with the initial $^{60}$Fe concentration
found in primitive type 3 ordinary chondrites \citep{Tac03}, lower than
initial values previously obtained for CAIs \citep{Bir88}, and supports the view
that $^{60}$Fe was likely not distributed homogeneously in the early solar
system. A more detailed H chondrite parent body modeling will be presented
elsewhere.

\section{Concluding remarks}

We constructed in this paper a model for the thermal evolution of the parent
body of H chondrites. The model considers compaction by cold pressing and
sintering by 'hot pressing'. The heat conductivity of the porous material was
determined by combining new data obtained by \citet{Kra11} for high porosity
material with data for porous chondrites \citep{Yom83}. A model was fitted to
data on the cooling history for two H chondrites, Kernouv\'e (H6) and Richardton
(H5), for which particular accurate data are available. It was shown that it is
possible to find a consistent fit for the parent body size and formation time 
that reproduces with good accuracy the empirically determined cooling history
of both, H5 and H6 chondrites.

For obtaining our consistent fit, it was necessary to include (besides radius of
the body and formation time) also the abundance of $^{60}$Fe into the
fitting procedure. A value of  $^{60}$Fe/$^{56}$Fe was determined, that is
within the range of values reported in the literature for different meteorites.
No other arbitrary fit parameters are required; all other properties of the
model are fixed by the physics of the problem.

The new model predicts rather shallow outer layers for petrologic types 3 to 5 
because of the strong blanketing effect of an outer powder layer, that escapes
sintering. In so far the model deviates considerable from previous models that
are based on the analytic model of \citep{Miy81}. Other properties of the model
are similar to older models; in particular radius and formation time are not
substantially different. 

The present model, though relaxing some earlier shortcomings, still has a
number of shortcomings. The most stringent is the instantaneous formation 
hypothesis, that needs to be relaxed because the formation time of the body by
run-away growth (of the order of 10$^ 5$a) is shorter than the decay time of
$^ {26\!}$Al, the main heating source, but probably not short enough for being
completely negligible. The second severe shortcoming is that heat conductivity
of porous media can not yet be treated from first principles on. This is not
possible with presently available computing resources, but this may change in
the near future. Other shortcomings are a rather simplistic treatment of
sintering and of the outer boundary temperature. We modelled in this
paper sintering with the same kind of theoretical description as \citet{Yom84}
in order that our results can be compared with that model. This treatment of
sintering should be replaced by more elaborated modern theories of hot pressing. 


\begin{acknowledgements}
We thank the referee S. Sahijpal for a constructive and helpful referee report.
This work was supported in part by `Forschergruppe 759' and in part by
`Schwerpunktprogramm 1385'. Both are supported by the `Deutsche 
Forschungs\-gemeinschaft (DFG)'. 
\end{acknowledgements}



\begin{thebibliography}{65}
\expandafter\ifx\csname natexlab\endcsname\relax\def\natexlab#1{#1}\fi

\bibitem[{Agee {et~al.}(1995)Agee, Li, Shannon, \& Circone}]{Age95}
Agee, C.~B., Li, J., Shannon, M.~C., \& Circone, S. 1995, \jgr, 100, 17725

\bibitem[{Akridge {et~al.}(1997)Akridge, Benoit, \& Sears}]{Akr97}
Akridge, G., Benoit, P.~H., \& Sears, D.~W.~G. 1997, Lunar Plan. Sc. Conf.,
  XXVIII, 1178

\bibitem[{{Akridge} {et~al.}(1998){Akridge}, {Benoit}, \& {Sears}}]{Akr98}
{Akridge}, G., {Benoit}, P.~H., \& {Sears}, D.~W.~G. 1998, \icarus, 132, 185

\bibitem[{Anders \& Grevesse(1989)}]{And89}
Anders, E. \& Grevesse, N. 1989, \gca, 53, 197

\bibitem[{Arzt(1982)}]{Arz82}
Arzt, E. 1982, Acta metall., 30, 1883

\bibitem[{Arzt {et~al.}(1983)Arzt, Ashby, \& Easterling}]{Arz83}
Arzt, E., Ashby, M.~F., \& Easterling, K.~E. 1983, Metallurgical Transact. A,
  14A, 211

\bibitem[{Barin(1995)}]{Bar95}
Barin, I. 1995, Thermochemical Data of Pure Substances, 3rd edn., Vol. I, II
  (VCH Verlagsgesellschaft Weinheim)

\bibitem[{Birck \& Lugmair(1988)}]{Bir88}
Birck, J.~L. \& Lugmair, G.~W. 1988, \epsl, 90, 131

\bibitem[{{Blum} {et~al.}(2006){Blum}, {Schr{\"a}pler}, {Davidsson}, \&
  {Trigo-Rodr{\'{\i}}guez}}]{Blu06}
{Blum}, J., {Schr{\"a}pler}, R., {Davidsson}, B.~J.~R., \&
  {Trigo-Rodr{\'{\i}}guez}, J.~M. 2006, \apj, 652, 1768

\bibitem[{Bouvier {et~al.}(2007)Bouvier, Blichert-Toft, Moynier, Vervoort, \&
  Albar\`ede}]{Bou07}
Bouvier, A., Blichert-Toft, J., Moynier, F., Vervoort, J., \& Albar\`ede, F.
  2007, \gca, 71, 1583

\bibitem[{{Bradley}(2010)}]{Bra03}
{Bradley}, J. 2010, in Lecture Notes in Physics, Vol. 609, Astromineralogy,
  2nd. Ed., ed. T.~K. {Henning} (Berlin: Springer), 259--276

\bibitem[{Finocchi \& Gail(1997)}]{Fin97}
Finocchi, F. \& Gail, H.-P. 1997, \aap, 327, 825

\bibitem[{Fischmeister \& Arzt(1983)}]{Fis83}
Fischmeister, H.~F. \& Arzt, E. 1983, Powder Metallurgy, 26, 82

\bibitem[{Fountain \& West(1970)}]{Fou70}
Fountain, J.~A. \& West, E.~A. 1970, \jgr, 75, 4063

\bibitem[{Ghosh \& McSween(1999)}]{Gho99}
Ghosh, A. \& McSween, H.~Y. 1999, \mps, 34, 121

\bibitem[{Ghosh {et~al.}(2003)Ghosh, Weidenschilling, \& McSween}]{Gho03}
Ghosh, A., Weidenschilling, S.~J., \& McSween, H.~Y. 2003, \mps, 38, 711

\bibitem[{Ghosh {et~al.}(2006)Ghosh, Weidenschilling, McSween, \&
  Rubin}]{Gho06}
Ghosh, A., Weidenschilling, S.~J., McSween, H.~Y., \& Rubin, A. 2006, in
  Meteorites and the Early Solar System II, ed. D.~S. Lauretta \& H.~Y.
  {McSween Jr.} (Tucson: Univ. of Arizona Press), 555--566

\bibitem[{G\"opel {et~al.}(1994)G\"opel, Manhes, \& Allegre}]{Goe94}
G\"opel, C., Manhes, G., \& Allegre, C.~J. 1994, \epsl, 121, 153

\bibitem[{Grimm \& McSween(1989)}]{Gri89}
Grimm, R.~E. \& McSween, H.~Y. 1989, \icarus, 82, 244

\bibitem[{Gupta \& Sahijpal(2010)}]{Gup10}
Gupta, G. \& Sahijpal, S. 2010, \jgr, 115, E08001

\bibitem[{{G{\"u}ttler} {et~al.}(2009){G{\"u}ttler}, {Krause}, {Geretshauser},
  {Speith}, \& {Blum}}]{Gue09}
{G{\"u}ttler}, C., {Krause}, M., {Geretshauser}, R.~J., {Speith}, R., \&
  {Blum}, J. 2009, \apj, 701, 130

\bibitem[{Harrison \& Grimm(2010)}]{Har10}
Harrison, K.~P. \& Grimm, R.~E. 2010, \gca, 74, 5410

\bibitem[{Herpfer {et~al.}(1994)Herpfer, Larimer, \& Goldstein}]{Her94}
Herpfer, M., Larimer, J., \& Goldstein, J. 1994, \gca, 58, 1353

\bibitem[{Hevey \& Sanders(2006)}]{Hev06}
Hevey, P.~J. \& Sanders, I.~S. 2006, \mps, 41, 95

\bibitem[{Hopfe \& Goldstein(2001)}]{Hop01}
Hopfe, W. \& Goldstein, J. 2001, \mps, 36, 135

\bibitem[{Jarosewich(1990)}]{Jar90}
Jarosewich, E. 1990, Meteoritics, 25, 323

\bibitem[{Jeager \& Nagel(1992)}]{Jae92}
Jeager, H.~M. \& Nagel, S.~R. 1992, Science, 255, 1523

\bibitem[{Kakar \& Chaklader(1967)}]{Kak67}
Kakar, A.~K. \& Chaklader, A.~C.~D. 1967, \jap, 38, 3223

\bibitem[{Kleine {et~al.}(2005)Kleine, Mezger, Palme, \& M\"unker}]{Kle05}
Kleine, T., Mezger, K., Palme, H. E.~S., \& M\"unker, C. 2005, \gca, 69, 5805

\bibitem[{Kleine {et~al.}(2008)Kleine, Touboul, {Van Orman}, Bourdon, Maden,
  Mezger, \& Halliday}]{Kle08}
Kleine, T., Touboul, M., {Van Orman}, J., {et~al.} 2008, \epsl, 270, 106

\bibitem[{{Korochantseva} {et~al.}(2007){Korochantseva}, {Trieloff}, {Lorenz},
  {Buykin}, {Ivanova}, {Schwarz}, {Hopp}, \& {Jessberger}}]{Kor07}
{Korochantseva}, E.~V., {Trieloff}, M., {Lorenz}, C.~A., {et~al.} 2007,
  Meteoritics and Planetary Science, 42, 113

\bibitem[{{Kothe} {et~al.}(2010){Kothe}, {G{\"u}ttler}, \& {Blum}}]{Kot10}
{Kothe}, S., {G{\"u}ttler}, C., \& {Blum}, J. 2010, \apj, 725, 1242

\bibitem[{Krause {et~al.}(2011{\natexlab{a}})Krause, Blum, Skorov, \&
  Trieloff}]{Kra11}
Krause, M., Blum, J., Skorov, Y., \& Trieloff, M. 2011{\natexlab{a}}, \icarus,
  214, 286

\bibitem[{Krause {et~al.}(2011{\natexlab{b}})Krause, Henke, Gail, Trieloff,
  Blum, Skorov, Schwarz, \& Kleine}]{Kra11b}
Krause, M., Henke, S., Gail, H.-P., {et~al.} 2011{\natexlab{b}}, \lpscl, 42,
  2696

\bibitem[{Larsson {et~al.}(1996)Larsson, Biwa, \& Stor{\aa}kers}]{Lar96}
Larsson, P.~L., Biwa, S., \& Stor{\aa}kers, B. 1996, Acta mater., 44, 3655

\bibitem[{Lodders {et~al.}(2009)Lodders, Palme, \& Gail}]{Lod09}
Lodders, K., Palme, H., \& Gail, H.~P. 2009, in Landolt-B\"ornstein, New
  Series, Group IV, Vol. 4, ed. J.~E. Tr\"umper (Berlin: Springer), 560--599

\bibitem[{McSween {et~al.}(2003)McSween, Ghosh, Grimm, Wilson, \&
  Young}]{McS03}
McSween, H., Ghosh, A., Grimm, R.~E., Wilson, L., \& Young, E.~D. 2003, in
  Asteroids III, ed. W.~F. Bottke (Tucson: Univ. of Arizona Press), 559--571

\bibitem[{Miyamoto {et~al.}(1981)Miyamoto, Fujii, \& Takeda}]{Miy81}
Miyamoto, M., Fujii, N., \& Takeda, H. 1981, \lpscl, 12B, 1145

\bibitem[{Nagasawa {et~al.}(2007)Nagasawa, Thommes, Kenyon, Bromley, \&
  Lin}]{Naga07}
Nagasawa, M., Thommes, E.~W., Kenyon, S.~J., Bromley, B.~C., \& Lin, D.~N.~C.
  2007, in Protostars and Planets V, ed. B.~Reipurt, D.~Jewitt, \& K.~Keil
  (Tucson: University of Arizona Press), 639--654

\bibitem[{Nyquist {et~al.}(2009)Nyquist, Kleine, Shih, \& Reese}]{Nyq09}
Nyquist, L.~E., Kleine, T., Shih, C.-Y., \& Reese, Y.~D. 2009, \gca, 73, 5115

\bibitem[{{Onoda} \& {Liniger}(1990)}]{Ono90}
{Onoda}, G.~Y. \& {Liniger}, E.~G. 1990, \prl, 64, 2727

\bibitem[{Pellas {et~al.}(1997)Pellas, Fi\'eni, Trieloff, \&
  Jessberger}]{Pel97}
Pellas, P., Fi\'eni, C., Trieloff, M., \& Jessberger, E. 1997, \gca, 61, 3477

\bibitem[{Quitt\'e {et~al.}(2007)Quitt\'e, Halliday, Meyer, Markowski,
  Latkoczy, \& G\"unther}]{Qui07}
Quitt\'e, G., Halliday, A.~N., Meyer, B.~S., {et~al.} 2007, \apj, 655, 678

\bibitem[{Quitt\'e {et~al.}(2010)Quitt\'e, Markowski, Latkoczy, Gabriel, \&
  Pack}]{Qui10}
Quitt\'e, G., Markowski, A., Latkoczy, C., Gabriel, A., \& Pack, A. 2010, \apj,
  720, 1215

\bibitem[{Rao \& Chaklader(1972)}]{Rao72}
Rao, A.~S. \& Chaklader, A.~C.~D. 1972, \jacs, 55, 596

\bibitem[{Rietmeijer(1993)}]{Rie93}
Rietmeijer, F. 1993, \epsl, 117, 609

\bibitem[{Rugel {et~al.}(2009)Rugel, Faestermann, Knie, Korschinek, Poutivtsev,
  Schumann, Kivel, G\"unther-Leopold, Weinreich, \& Wohlmuther}]{Rug09}
Rugel, G., Faestermann, T., Knie, K., {et~al.} 2009, \prl, 103, 072502

\bibitem[{Sahijpal {et~al.}(2007)Sahijpal, Soni, \& Gupta}]{Sah07}
Sahijpal, S., Soni, P., \& Gupta, G. 2007, \mps, 42, 1529

\bibitem[{Schwenn \& Goetze(1978)}]{Sch78}
Schwenn, M.~B. \& Goetze, C. 1978, \tectp, 48, 41

\bibitem[{Scott(2007)}]{Sco97}
Scott, E.~R.~D. 2007, \areps, 35, 577

\bibitem[{Scott(1962)}]{Sco62}
Scott, G.~D. 1962, \nat, 194, 956

\bibitem[{Senshu(2004)}]{Sen04}
Senshu, H. 2004, Lunar Plan. Sc. Conf., XXXV, 1557

\bibitem[{Stor{\aa}kers {et~al.}(1999)Stor{\aa}kers, Fleck, \&
  McMeeking}]{Sto99}
Stor{\aa}kers, B., Fleck, N.~A., \& McMeeking, R.~M. 1999, J. Mech. Phys.
  Solids, 47, 785

\bibitem[{{Tachibana} \& {Huss}(2003)}]{Tac03}
{Tachibana}, S. \& {Huss}, G.~R. 2003, \apjl, 588, L41

\bibitem[{Taylor {et~al.}(1987)Taylor, Maggiore, Scott, \& Keil}]{Tay87}
Taylor, G.~J., Maggiore, P., Scott, E.~R.~D., \& Keil, A.~E. 1987, \icarus, 69,
  1

\bibitem[{Trieloff {et~al.}(2003)Trieloff, Jessberger, Herrwerth, Hopp,
  Fi\'eni, Gh\'elis, Bourot-Denise, \& Pellas}]{Tri03}
Trieloff, M., Jessberger, E., Herrwerth, I., {et~al.} 2003, \nat, 422, 502

\bibitem[{{Van Schmus}(1995)}]{VSc95}
{Van Schmus}, W.~R. 1995, in Global Earth Physics. A Handbook on Physical
  Constants. AGU Reference Shelf 1 (American Geophysical Union), 283--291

\bibitem[{{Wehrstedt} \& {Gail}(2002)}]{Weh02}
{Wehrstedt}, M. \& {Gail}, H. 2002, \aap, 385, 181

\bibitem[{{Wehrstedt} \& {Gail}(2008)}]{Weh08}
{Wehrstedt}, M. \& {Gail}, H. 2008, ArXiv e-prints, 0804.3377

\bibitem[{Weidenschilling \& Cuzzi(2006)}]{Wei06}
Weidenschilling, S.~J. \& Cuzzi, J.~N. 2006, in {Meteorites and the Early Solar
  System II}, ed. D.~S. Lauretta \& H.~Y. McSween (Tucson: University of
  Arizona Press), 473--485

\bibitem[{Weidling {et~al.}(2009)Weidling, G\"uttler, Blum, \& Brauer}]{Wei09}
Weidling, R., G\"uttler, C., Blum, J., \& Brauer, F. 2009, \apj, 696, 2036

\bibitem[{Wilkinson \& Ashby(1975)}]{Wil75}
Wilkinson, D. \& Ashby, M. 1975, Acta Metallurgica, 23, 1277

\bibitem[{Yang {et~al.}(2000)Yang, Zou, \& Yu}]{Yan00}
Yang, R.~Y., Zou, R.~P., \& Yu, A.~B. 2000, \pre, 62, 3900

\bibitem[{Yomogida \& Matsui(1983)}]{Yom83}
Yomogida, K. \& Matsui, T. 1983, \jgr, 88, 9513

\bibitem[{Yomogida \& Matsui(1984)}]{Yom84}
Yomogida, K. \& Matsui, T. 1984, Earth \& Plan. Sci. L., 68, 34

\end{thebibliography}
\end{document}